\documentclass[aps,prb,twocolumn,nofootinbib,longbibliography]{revtex4-2}

\usepackage{microtype} % improved kerning/symbol stretching

\usepackage{amsmath}
\usepackage{amsfonts}
\usepackage{graphicx}
\usepackage{booktabs}
\usepackage{braket}

\usepackage{lipsum}
\usepackage{mwe}
\usepackage{caption}
\usepackage{subcaption}
\usepackage{ntheorem}

\usepackage{multirow}
\usepackage{xcolor}
\usepackage{hyperref}
\usepackage{bbm}
\usepackage{bm}

\hypersetup{colorlinks,
	linkcolor={blue!75!black!80!yellow},
	citecolor={blue!75!black!80!yellow},
	urlcolor={blue!75!black!80!yellow}
}

\begin{document}
\begin{abstract}

The decomposition of the Hilbert space of a quantum many-body system into the irreducible representations of its bond and commutant algebras yields a finer structure of dynamically isolated subspaces than the mere decomposition into symmetry sectors. While it has 
been recognized that subspaces associated with low bond-irrep dimensions $D_{\lambda}$ tend to violate the 
eigenstate thermalization hypothesis (ETH), here we show that $D_\lambda$ controls thermalization continuously across the full spectrum of dynamical subspaces. Using 
SU(2)-symmetric spin-$1/2$ chains as a paradigmatic example, we 
demonstrate that $\log D_\lambda$ quantitatively accounts for the 
average eigenstate entanglement entropy within each sector, establishing 
a thermalization hierarchy that interpolates from exact quantum many-body 
scars at $D_\lambda = 1$ to volume-law ergodic states at large 
$D_\lambda$. To make this concrete, we introduce the notion of 
\emph{irreducible degrees of freedom} (IDOF), defined as the number of 
independently-varying spatial coordinates parametrizing a many-body 
state within a given bond-algebra sector, which provides a microscopic 
interpretation of $D_\lambda$ and of the resulting thermalization 
hierarchy. Finally, we show that by selectively breaking symmetries 
while preserving chosen bond-algebra sectors, one can embed families of 
nonthermal eigenstates at prescribed entanglement levels into an 
otherwise ergodic spectrum, generalizing restricted spectrum-generating 
algebras from towers of individual states to entire dynamical 
subspaces.
\end{abstract}

\preprint{APS/123-QED}
\author{Pedro Fittipaldi de Castro}
\author{Wladimir A. Benalcazar}
\email{benalcazar@emory.edu}
\affiliation{
Department of Physics, Emory University, Atlanta, GA 30322, USA
}
\title{Thermalization hierarchy from irreducible degrees of freedom} % Force line breaks with \\
\maketitle

\begin{figure*}[t]
    \centering
    \includegraphics[width=0.9\textwidth]{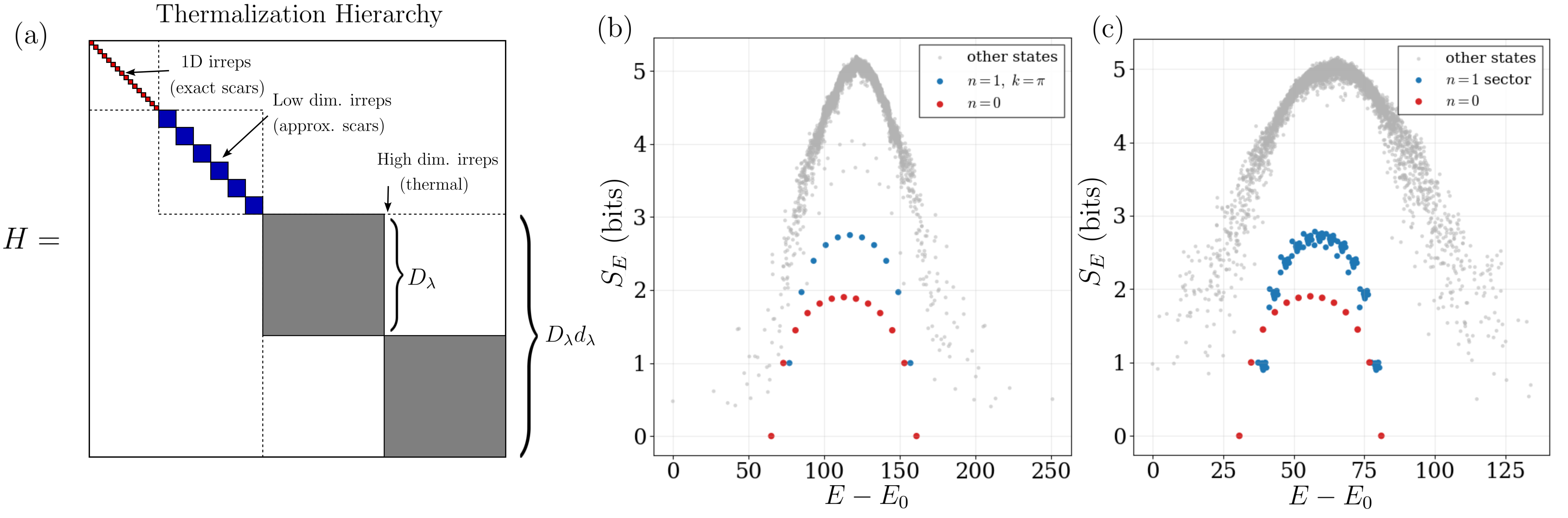}
    \caption{Thermalization hierarchy. (a) Decomposition of the Hamiltonian \eqref{LocalHam} in the basis \eqref{SpecialBasis}, showing dynamically isolated subspaces as diagonal blocks. Half-chain entanglement entropy of the eigenstates of (b) the local, translation invariant Hamiltonian \eqref{HamiltonianOpi} with $\kappa=2.5$ and of (c) the nonlocal, disordered Hamiltonian~\eqref{HNonlocal} with $\kappa=0.02$. Calculations done under periodic boundary conditions with parameters $J_{1}=J_{2}=J_{3}=1$, and $B=8$. Energies are shown relative to $E_{\mathrm{0}}$ of the ferromagnetic state $\ket{\uparrow,\uparrow,\dots,\uparrow}$.}
    \label{HamiltonianDecomposition}
\end{figure*}

\section{Introduction}

The eigenstate thermalization hypothesis (ETH) is the cornerstone of our understanding of thermalization in isolated quantum many-body systems. In its standard formulation, ETH states that energy eigenstates in the bulk of the spectrum reproduce thermal expectation values of few-body observables within resolved symmetry sectors~\cite{deutsch2018eigenstate,PhysRevE.50.888,d2016quantum,rigol2008thermalization,mori2018thermalization,nandkishore2015many}. Such thermal eigenstates typically exhibit volume-law entanglement entropy.

The validity of ETH is widely regarded as the characteristic signature of quantum chaos in many-body systems, and its failure as the hallmark of non-ergodic behavior. Prominent examples of ETH violation include many-body localization (MBL)~\cite{PhysRev.109.1492,BASKO20061126,PhysRevLett.95.206603,PhysRevB.75.155111,PhysRevB.82.174411,PhysRevLett.111.127201,PhysRevB.90.174202,nandkishore2015many,ALET2018498}, quantum many-body scars (QMBS)~\cite{turner2018weak,PhysRevB.98.155134,bernien2017probing,PhysRevLett.119.030601,PhysRevLett.122.040603,PhysRevB.101.174308}, and Hilbert space fragmentation ~\cite{PhysRevX.10.011047,moudgalya2022quantum,PhysRevB.101.174204,PhysRevB.101.125126,Moudgalya_2021,scherg2021observing}. While these phenomena were initially identified through distinct mechanisms, a unifying algebraic framework has emerged in recent years. The bond algebra of a Hamiltonian, i.e., the operator algebra generated by its local interaction terms, and its commutant algebra together determine, via the double commutant theorem, a fine-grained decomposition of the Hilbert space into irreducible subspaces that can be strictly smaller than the symmetry sectors imposed by global conserved charges ~\cite{MOUDGALYA2023169384}. Scars and fragmentation are naturally accommodated within this framework as consequences of a nontrivial bond algebra structure, rather than as anomalies requiring separate explanations~\cite{PhysRevX.14.041069,PhysRevX.12.011050,PhysRevLett.126.120604}.

In this paper we argue that this algebraic decomposition is not merely a diagnostic for ETH violation, but identifies the correct subspaces within which ETH itself should be formulated. For finite-dimensional quantum spin systems in which the bond algebra acts reducibly on one or more symmetry sectors, the standard symmetry-resolved statement of ETH is too coarse: thermalization, when it occurs, does so within individual bond algebra irreps, not within entire symmetry sectors. The dimensions of these irreps may vary widely, from exponentially large subspaces where the statistical conditions for ETH are robustly satisfied, to polynomially large subspaces where thermalization is suppressed, down to one-dimensional irreps that constitute exact QMBS, as we illustrate schematically in Fig.~\ref{HamiltonianDecomposition}(a). Using spin-1/2 chains as a paradigmatic example, we make this hierarchy precise by introducing the notion of irreducible degrees of freedom, \emph{i.e.} the number of independent coordinates parametrizing a many-body state transforming according to a given bond irrep, as a quantitative control parameter for thermalization.  As we show, the logarithm of the irrep dimension quantitatively accounts for the eigenstate entanglement entropy within each subspace, confirming that irreducible degrees of freedom provide a sharp measure of the thermalization capacity available within a given bond algebra sector. This reveals a thermalization hierarchy of which the familiar dichotomy between thermalizing and non-ergodic behavior is only the coarsest expression.

Finally, we demonstrate that the thermalization hierarchy established by the bond algebra decomposition is not merely a diagnostic but a constructive framework. By identifying the conditions under which a restricted double-commutant decomposition organizes selected bond algebra irreps into towers of protected subspaces, we obtain a systematic prescription for embedding QMBS and other subthermal states of varying levels of entanglement and dynamical complexity into an otherwise thermal spectrum, as shown in Fig.~\ref{HamiltonianDecomposition}(b,c). This generalizes the notion of a restricted spectrum-generating algebra~\cite{PhysRevD.2.2944,bohm2024twenty,PhysRevB.102.085140},which emerges as the special case in which the protected subspace is one-dimensional in the commutant sector. The resulting picture helps to unify the phenomenology of QMBS, Hilbert space fragmentation, and partial thermalization under a single algebraic principle, and provides a practical toolkit for the design of non-ergodic behavior in quantum many-body systems.

\section{Thermalization Hierarchy from bond-algebra irreps}

Let us briefly introduce the notions of bond and commutant algebras in the language of Refs.~\cite{MOUDGALYA2023169384,PhysRevX.14.041069,PhysRevX.12.011050}. Consider Hamiltonians of the form
\begin{equation}
    \hat{H} = \sum_{\alpha} J_{\alpha} \hat{H}_{\alpha}, 
    \label{LocalHam}
\end{equation}
where $\hat{H}_{\alpha}$ are either strictly local operators (with support over a few sites of a lattice) or sums of such operators acting on a Hilbert space $\mathcal{H}$, and $J_{\alpha} \in \mathbb{R}$ are arbitrary coefficients. The bond algebra~\footnote{Refs.~\cite{MOUDGALYA2023169384,PhysRevX.14.041069,PhysRevX.12.011050} distinguish bond algebras as being generated by strictly local (or sums of strictly local) operators. For simplicity, we use the term ``bond'' throughout.} $\mathcal{A}_{\mathrm{bond}}$ associated with Hamiltonians of the form \eqref{LocalHam} consists of the identity $\hat{\mathbb{I}}$, all $\hat{H}_{\alpha}$, and all of their possible products and linear combinations. That is, for all $ \hat{H}_{\alpha},\hat{H}_{\beta} \in \mathcal{A}_{\mathrm{bond}}$ and $c_{\alpha},c_{\beta} \in \mathbb{C}$, we have
\begin{align}
    &\hat{H}_{\alpha}\hat{H}_{\beta} \in \mathcal{A}_{\mathrm{bond}} \label{A1} \\ 
    &c_{\alpha}\hat{H}_{\alpha}+c_{\beta}\hat{H}_{\beta} \in \mathcal{A}_{\mathrm{bond}}. \nonumber
\end{align}
From now on, we denote the associative algebra generated by operators $\hat{H}_{\alpha}$ according to properties \eqref{A1} as $\mathcal{A}_{\mathrm{bond}}=\mathrm{Alg}\{\hat{H}_{\alpha}\}$.

To every bond algebra, there is an associated commutant algebra $\mathcal{A}_{\mathrm{comm}}$ formed by all the operators $\hat{O}_{\alpha}$ that commute with every single term in the Hamiltonian \eqref{LocalHam}, \emph{i.e.},
\begin{equation}
    [\hat{H}_{\alpha}, \hat{O}_{\beta}] = 0 \quad \forall \ \alpha,\beta.
    \label{CommutantAlgebra}
\end{equation}
Note that the condition above is stronger than commuting with $\hat{H}$ itself, and thus $\mathcal{A}_{\mathrm{comm}}$ is generally a subset of the conventional symmetries of \eqref{LocalHam}. Meanwhile, any product $\hat{O}_{\alpha}\hat{O}_{\beta}$ or linear combination $c_{\alpha}\hat{O}_{\alpha}+c_{\beta}\hat{O}_{\beta}$ of operators satisfying \eqref{CommutantAlgebra} also commutes with every term in the Hamiltonian \eqref{LocalHam}. Therefore, $\mathcal{A}_{\mathrm{comm}}= \mathrm{Alg}\{\hat{O}_{\alpha}\}$ is closed under the same two operations as $\mathcal{A}_{\mathrm{bond}}$.

If $\hat{H}_{\alpha}^{\dagger} = \hat{H}_{\alpha}$ for all $\alpha$, both associative algebras
$\mathcal{A}_{\mathrm{bond}}$ and $\mathcal{A}_{\mathrm{comm}}$ are closed under Hermitian conjugation~\cite{MOUDGALYA2023169384}. 
In finite dimensions, they form $\dagger$-algebras and therefore coincide with their double commutants. 
As a result, $\mathcal{A}_{\mathrm{bond}}$ and $\mathcal{A}_{\mathrm{comm}}$ are mutual commutants (centralizers)~\cite{landsman1998lecture,harlow2017ryu,kabernik2021reductions}, 
and the Hilbert space admits the decomposition
\begin{equation}
    \mathcal{H} = \bigoplus_{\lambda \in \Lambda} \left( \mathcal{H}^{\mathrm{bond}}_{\lambda} \otimes \mathcal{H}^{\mathrm{comm}}_{\lambda} \right).
    \label{Decomposition}
\end{equation}
In the expression above, $\lambda$ labels sectors of the Hilbert space associated with the irreps of $\mathcal{A}_{\mathrm{bond}}$ and $\mathcal{A}_{\mathrm{comm}}$, and $\Lambda$ is the set of all possible irreps. Concretely, there exists an orthonormal basis adapted to this decomposition,
\begin{align}
    \ket{w_{\lambda,\alpha,\beta}}=\ket{u_{\lambda,\alpha}} \otimes \ket{v_{\lambda,\beta}}, \label{SpecialBasis}
\end{align}
with $\alpha = 1,\dots,D_{\lambda}$ and $\beta = 1,\dots,d_{\lambda}$, such that any operator 
$\hat{O}_{\mathrm{bond}} \in \mathcal{A}_{\mathrm{bond}}$ acts as
\begin{equation}
    \hat{O}_{\mathrm{bond}}(\ket{u_{\lambda,\alpha}} \otimes \ket{v_{\lambda,\beta}})
    =
    \sum_{\alpha'=1}^{D_{\lambda}}
    O_{\mathrm{bond},\lambda}^{\alpha\alpha'}
    \ket{u_{\lambda,\alpha'}} \otimes \ket{v_{\lambda,\beta}}, \label{BlockDiagonalization}
\end{equation}
while any $\hat{O}_{\mathrm{comm}} \in \mathcal{A}_{\mathrm{comm}}$ acts as
\begin{equation}
    \hat{O}_{\mathrm{comm}}(\ket{u_{\lambda,\alpha}} \otimes \ket{v_{\lambda,\beta}})
    =
    \ket{u_{\lambda,\alpha}} \otimes
    \sum_{\beta'=1}^{d_{\lambda}}
    O_{\mathrm{comm},\lambda}^{\beta\beta'}
    \ket{v_{\lambda,\beta'}}.
\end{equation}
Therefore, $\mathcal{H}^{\mathrm{bond}}_{\lambda}$ ($\mathcal{H}^{\mathrm{comm}}_{\lambda}$) carries a $D_{\lambda}$-dimensional ($d_{\lambda}$-dimensional) irrep of $\mathcal{A}_{\mathrm{bond}}$ ($\mathcal{A}_{\mathrm{comm}}$). As a result, the dimension of each Hilbert space sector $\mathcal{H}^{\mathrm{bond}}_{\lambda} \otimes \mathcal{H}^{\mathrm{comm}}_{\lambda}$ in \eqref{Decomposition} is $D_{\lambda}d_{\lambda}$. Running over all the allowed irreps, we must exhaust the full Hilbert space, $\sum_{\lambda}D_{\lambda}d_{\lambda}=\mathrm{dim} \ \mathcal{H}$, as illustrated schematically in Fig.~\ref{HamiltonianDecomposition}(a).

In particular, by the properties \eqref{A1}, the Hamiltonian \eqref{LocalHam} is an element of $\mathcal{A}_{\mathrm{bond}}$ and can be block diagonalized as in \eqref{BlockDiagonalization}. The blocks
\begin{equation}
    H_{\lambda} = \sum_{\alpha,\alpha'}\ket{u_{\lambda,\alpha'}}H_{\lambda}^{\alpha\alpha'}\bra{u_{\lambda,\alpha}} \otimes \mathbb{I}_{d_{\lambda}}~\footnote{Here, $\mathbb{I}_{d_{\lambda}}=\sum_{\beta}\ket{v_{\lambda,\beta}}\bra{v_{\lambda,\beta}}$} \label{DecompositionHamiltonian}
\end{equation}
act on a sector of dimension $D_{\lambda} d_{\lambda}$ and can be further decomposed into $d_{\lambda}$ copies of smaller $D_{\lambda}$-dimensional blocks on which $H_{\lambda}^{\alpha\alpha'}$ acts nontrivially. These smaller Blocks are dynamically isolated and are naturally associated with Krylov subspaces~\cite{liesen2013krylov,moudgalya2022quantum}. In the context of ETH violation, sectors with small $D_{\lambda}$ are of special interest. When $D_{\lambda} = 1$, the Hamiltonian reduces to a scalar within each block, yielding a $d_{\lambda}$-fold degenerate eigenspace spanned by states satisfying $\hat{H}\ket{\psi} = E_{\lambda}\ket{\psi}$ [red blocks in Fig~\ref{HamiltonianDecomposition}(a)]. We refer to these as ``bond singlets'' of the algebra $\mathcal{A}_{\mathrm{bond}}$. Importantly, this fact does not rely on integrability.
Even when the higher-dimensional blocks cannot be diagonalized analytically, the one-dimensional blocks yield exact eigenstates that are fixed entirely by the algebraic structure of $\mathcal{A}_{\mathrm{bond}}$. As such, one can naturally interpret them as QMBSs~\cite{moudgalya2022quantum,PhysRevLett.125.230602,PhysRevResearch.3.043156}.

Even in generic nonintegrable regimes where no closed-form expressions for the energy eigenstates within Krylov subspaces with $D_{\lambda}>1$ are available, the states in each such invariant subspace may still be interpreted as ``bond multiplets'' associated with an irreducible representation of the bond algebra. In this sense, the decomposition of $\mathcal{A}_{\mathrm{bond}}$ into irreducible sectors of dimensions
\begin{equation}
    D_{\lambda_1}\leq D_{\lambda_2}\leq \cdots \leq D_{\lambda_{\mathrm{max}}} \label{ThermalizationHierarchy}
\end{equation}
naturally defines a thermalization hierarchy in which low-dimensional irreps tend to exhibit subthermal behavior [blue blocks in Fig.~\ref{HamiltonianDecomposition}(a)], while sufficiently large irreps [gray blocks in Fig.~\ref{HamiltonianDecomposition}(a)] are expected to satisfy the Eigenstate Thermalization Hypothesis (ETH).

We interpret this hierarchy in terms of irreducible degrees of freedom (IDOF) associated with the bond multiplets, which provide a more concrete microscopic picture of how thermalization emerges in a given many-body system. In particular, low-dimensional irreps correspond to collective dynamics involving only a few effective degrees of freedom, whereas larger irreps support progressively richer internal dynamics consistent with ergodic thermalization. While the underlying framework is considerably more general, this hierarchy becomes especially transparent in nonintegrable spin-$1/2$ systems with conventional or dynamical SU(2) symmetry, where the bond multiplets acquire a natural interpretation in terms of collective spin dynamics, as we now investigate in detail.

\section{Irreducible degrees of freedom in the spin-1/2 chain}

As a simple example, which is also featured in Refs.~\cite{MOUDGALYA2023169384,PhysRevX.14.041069} in the context of bond singlets, consider the family of SU(2)-symmetric Heisenberg Hamiltonians
\begin{equation}
\hat{H}_{\mathrm{SU(2)}} =
\sum_{i,j} J_{ij}
\hat{\bm S}_{i}\cdot\hat{\bm S}_{j},
\label{Heinsenberg}
\end{equation}
where
$\hat{\bm S}_{i}=(\hat S^x_i,\hat S^y_i,\hat S^z_i)$
are spin-$1/2$ operators acting on site
$i\in{0,1,\dots,L-1}$. The exchange interactions
$\hat{\bm S}_{i}\cdot\hat{\bm S}_{j}$ are isotropic, meaning that
$\hat{H}_{\mathrm{SU(2)}}$ is invariant under rigid rotations of all spins,
i.e., under the action of global SU(2). These rotations are implemented in
the Hilbert space by the unitary operators $
U(\hat{\bm n},\theta)
=
e^{-\mathrm{i}\theta\hat{\bm n}\cdot\hat{\bm S}_{\mathrm{tot}}},
$
where $\hat{\bm n}$ is an arbitrary direction in three dimensions and
$\hat{\bm S}_{\mathrm{tot}}=\sum_{j}\hat{\bm{S}}_j$ is the total spin operator. In particular, rotations $U_z(\theta)=e^{-i\theta \hat S^z_{\mathrm{tot}}}$ about a fixed axis form a U(1) subgroup of SU(2). Invariance under this subgroup is equivalent
to the conservation of the total magnetization $\hat S^z_{\mathrm{tot}}$.

\subsection{Bond algebra}

The bond algebra generated by the two-body operators
$\hat{\bm S}_{i}\cdot\hat{\bm S}_{j}$
is equivalent to the algebra generated by pairwise spin permutations. To see this, let
\begin{equation}
    \ket{\sigma_0,\sigma_1,\dots,\sigma_{L-1}},
    \qquad
    \sigma_i\in\{\uparrow,\downarrow\},
\end{equation}
denote an orthonormal basis of the Hilbert space, where
\begin{equation}
    \hat S_i^z\ket{\uparrow}_i
    =
    +\frac12\ket{\uparrow}_i,
    \qquad
    \hat S_i^z\ket{\downarrow}_i
    =
    -\frac12\ket{\downarrow}_i.
\end{equation}
Now define the operators
\begin{equation}
    \hat{\pi}_{ij}
    =
    2\hat{\bm S}_{i}\cdot\hat{\bm S}_{j}
    +
    \frac12\hat{\mathbb I}.
\end{equation}
These act as spin-exchange operators,
\begin{equation}
    \hat{\pi}_{ij}
    \ket{\cdots,\sigma_i,\cdots,\sigma_j,\cdots}
    =
    \ket{\cdots,\sigma_j,\cdots,\sigma_i,\cdots},
\end{equation}
thereby implementing pairwise permutations of spins. Consequently,
\begin{equation}
    \mathcal A_{\mathrm{bond}}
    =
    \mathrm{Alg}\{\hat{\bm S}_{i}\cdot\hat{\bm S}_{j}\}
    =
    \mathrm{Alg}\{\hat{\pi}_{ij}\}
    =
    \mathbb C[S_L], \label{BondAlgebraSpins}
\end{equation}
namely the algebra generated by the group of permutations (also known as the symmetric group) $S_L$ acting on the $L$ lattice sites.

\subsection{Commutant algebra}

Since $\hat{\pi}_{ij}$ merely exchanges the spin operators on sites $i$ and $j$, leaving the sum
$\hat{\bm S}_{\mathrm{tot}}=\sum_j \hat{\bm S}_j$ invariant, we conclude that the total spin operators
\begin{equation}
    \hat{S}^{\alpha}_{\mathrm{tot}} = \sum_{j=0}^{L-1}\hat{S}^{\alpha}_{j}, \quad \text{for} \ \alpha \in \{x,y,z,+,-\}, \label{TotalSpins}
\end{equation}
satisfy $[\hat{\pi}_{ij},\hat{S}^{\alpha}_{\mathrm{tot}}]=0$, and thus commute with all elements of the bond algebra. They generate the commutant algebra
\begin{equation}
    \mathcal{A}_{\mathrm{comm}} = \mathrm{Alg}\{\hat{S}^{x}_{\mathrm{tot}},\hat{S}^{y}_{\mathrm{tot}},\hat{S}^{z}_{\mathrm{tot}}\} = U[\mathfrak{su}(2)],
\end{equation}
known as the universal enveloping algebra of the Lie algebra $\mathfrak{su}(2)$~\cite{PhysRevX.12.011050}.

\begin{figure}
    \centering
    \includegraphics[width=\linewidth]{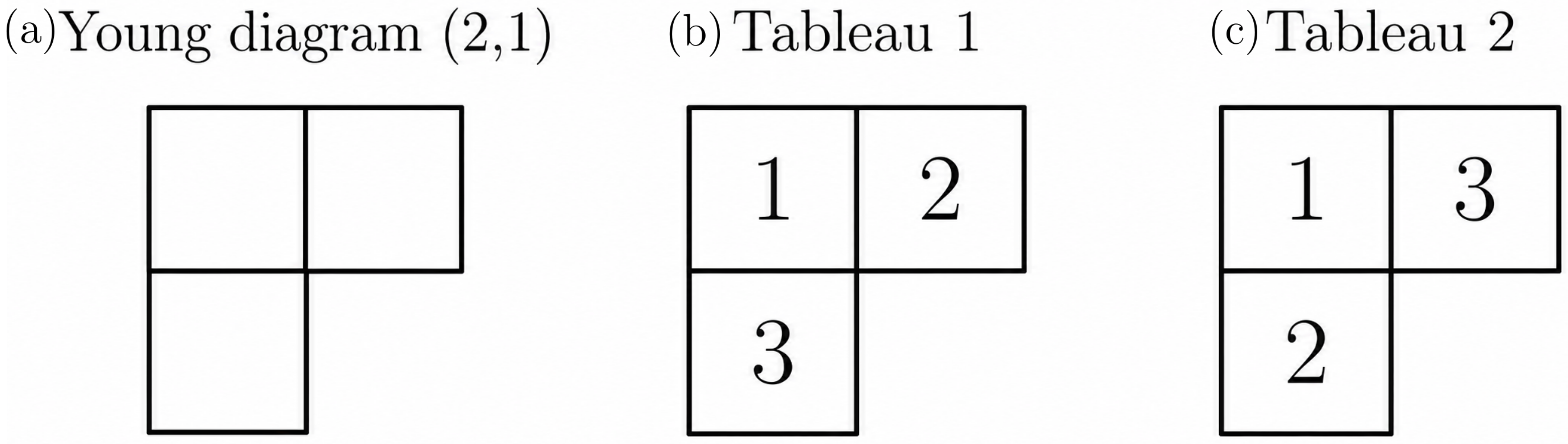}
    \caption{Permutation irreps as Young diagrams. (a) Diagram corresponding to the $(2,1)$ irrep of $S_3$. (b) and (c) are its two Young tableaux, which yield a two-dimensional irrep.}
    \label{Tableaux}
\end{figure}

\begin{figure*}[t]
    \centering
    \includegraphics[width=0.88\linewidth]{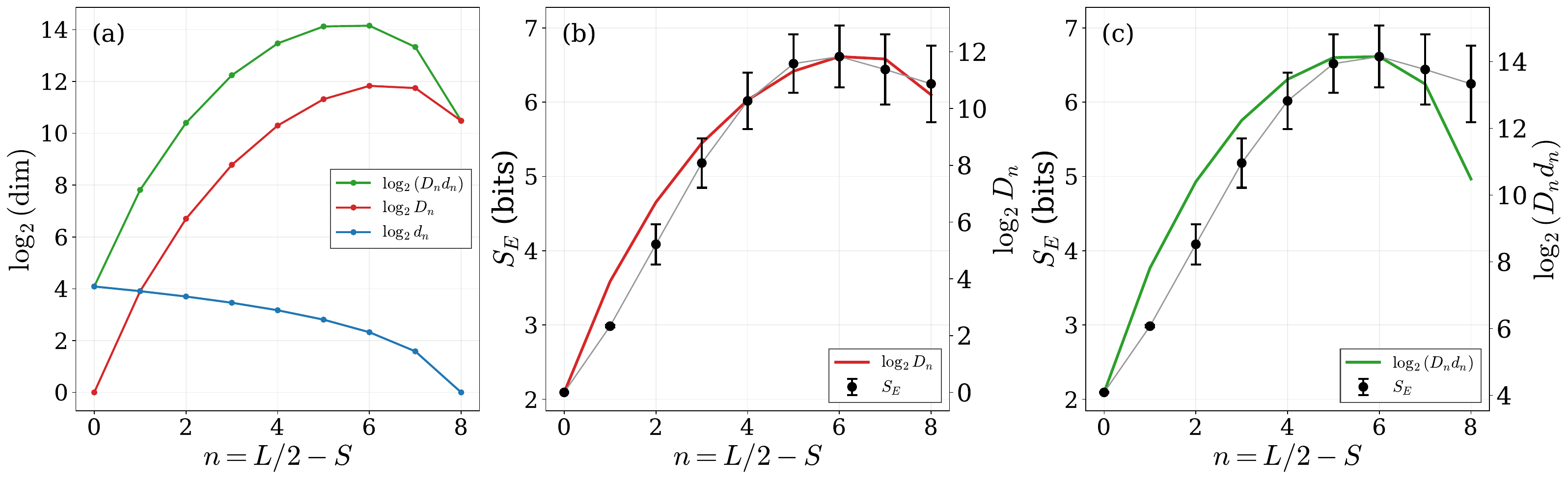}
    \caption{(a) Dimensions $D_{n}$ ($d_n$) of the bond (commutant) algebra irreps as well as the dimension $D_nd_n$ of the total spin sector versus $n=L/2-S$ for $L=16$. Half-chain entanglement entropy $S_E$ averaged over all energy eigenstates of \eqref{Heinsenberg} with a given $n$ and its comparison with the logarithm of (b) the bond-irrep dimensions $D_n$ and (c) the spin sector dimensions $D_nd_n$.}
    \label{EffectiveDimension}
\end{figure*}

\subsection{Schur-Weyl duality and irreps of the bond and commutant algebras}

The key to identifying the sectors that partition the Hilbert space as in Eq.~\eqref{Decomposition} is to determine the common center $
    \mathcal Z
    =
    \mathcal A_{\mathrm{bond}}
    \cap
    \mathcal A_{\mathrm{comm}}$, the set of operators commuting with all elements of $\mathcal A_{\mathrm{bond}}$ and $\mathcal A_{\mathrm{comm}}$. The identity operator $\hat{\mathbb I}$ is trivially always an element of $\mathcal Z$.

In the present case, the total-spin operator
\begin{equation}
    \hat{\bm S}_{\mathrm{tot}}^{2}
    =
    (\hat S^{x}_{\mathrm{tot}})^2
    +
    (\hat S^{y}_{\mathrm{tot}})^2
    +
    (\hat S^{z}_{\mathrm{tot}})^2
\end{equation}
also belongs to the common center $\mathcal Z$. Indeed, $\hat{\bm S}_{\mathrm{tot}}^{2}$ is an element of $\mathcal A_{\mathrm{comm}}$ and satisfies
\begin{equation}
    [\hat{\bm S}_{\mathrm{tot}}^{2},\hat S^\alpha_{\mathrm{tot}}]=0,
    \qquad
    \alpha\in\{x,y,z,+,-\}.
\end{equation}
Hence, it commutes with all elements of the commutant algebra, so it also belongs to $\mathcal A_{\mathrm{bond}}$. The total-spin Casimir $\hat{\bm S}_{\mathrm{tot}}^{2}$ has eigenvalues
\begin{equation}
    S(S+1),
    \qquad
    S\in
    \begin{cases}
        \{0,1,\dots,\frac{L}{2}\}, & L \ \text{even}, \\
        \{\frac12,\frac32,\dots,\frac{L}{2}\}, & L \ \text{odd},
    \end{cases}
\end{equation}
with associated orthogonal projectors $\hat P_S$ onto the total-spin-$S$ sectors. These projectors partition the Hilbert space as
\begin{equation}
    \mathcal H
    =
    \bigoplus_S
    \mathcal{H}_{S} = \bigoplus_S
    \left(\mathcal{H}_{S}^{\mathrm{bond}}\otimes \mathcal{H}^{\mathrm{comm}}_{S}\right), \label{SchurWeyl}
\end{equation}
where each spin sector can be further split into spaces $\mathcal{H}_{S}^{\mathrm{bond}}$ and $\mathcal{H}_{S}^{\mathrm{comm}}$ which carry irreps of $\mathcal A_{\mathrm{bond}}$ and $\mathcal A_{\mathrm{comm}}$. This particular form of the decomposition \eqref{Decomposition} is known as the Schur-Weyl duality~\cite{fulton2013representation,stevens2016schur}.

The irreducible representations of the symmetric group $S_L$ are labeled by partitions of $L$~\cite{james2006representation,fulton2013representation}. A partition is simply a way of writing $L$ as a sum of positive integers in nonincreasing order. For example, for $S_3$, the possible partitions are $
3$, $2+1$ and $1+1+1$,
which we denote by $
(3)$, $(2,1)$ and $(1,1,1)$.

Each partition can be represented visually by a Young diagram [see Fig. \ref{Tableaux}(a)]. This is a diagram made of boxes, where the number of boxes in each row is given by the corresponding part of the partition. Thus $(3)$ is represented by one row of three boxes, $(2,1)$ by one row of two boxes and a second row of one box, and $(1,1,1)$ by three rows of one box each.

The dimension of the irrep of $S_L$ associated with a Young diagram is obtained by counting the number of ways of filling the boxes with the labels $(1,\dots,L)$, provided each label appears exactly once, and the numbers increase from left to right along each row and from top to bottom along each column. For $S_3$, the diagrams $(3)$ and $(1,1,1)$ each have only one way of being filled, so their representations are one-dimensional. Meanwhile, the diagram $(2,1)$ has two possible fillings, as Figs. \ref{Tableaux} (b,c) show, so its representation is two-dimensional.

In the Schur--Weyl decomposition \eqref{SchurWeyl}, each total-spin sector $S$ is associated with a two-row Young diagram $(L-n,n)$, where
\begin{equation}
    n = L/2 - S \in \left\{0,1,\dots,\left\lfloor \frac{L}{2} \right\rfloor\right\}. \label{definitionn}
\end{equation}
The dimension of the corresponding irrep of $S_L$ can be obtained directly from the tableau-counting rule above, or equivalently from the hook length formula~\cite{fulton2013representation,stevens2016schur}, giving
\begin{equation}
    \dim \mathcal{H}_S^{\mathrm{bond}} = D_n = \binom{L}{n} - \binom{L}{n-1}. \label{DS}
\end{equation}

Meanwhile, the irrep of $\mathfrak{su}(2)$ with total-spin $S$ has dimension
\begin{equation}
    \dim \mathcal{H}_S^{\mathrm{comm}}= d_{n} = 
    L-2n+1. \label{ds}
\end{equation}

Fig.~\ref{EffectiveDimension}(a) shows the bond-irrep dimension $D_n$ (red), the commutant multiplicity $d_n$ (blue), and the total-spin-sector dimension $D_n d_n$ (green) as functions of $n$ for a finite system of size $L=16$.\footnote{Although $D_n$ and $D_n d_n$ exhibit qualitatively similar scaling behaviors, their maxima occur at different values of $n$.} According to expression \eqref{ThermalizationHierarchy}, the degree of ergodic mixing of energy eigenstates should be controlled by the bond-irrep dimension $D_n$. In particular, the logarithm of $D_n$ is extensive in the thermodynamic limit, and we interpret it as an effective entropy characterizing the dynamically isolated subspace associated with a given bond irrep. Figures~\ref{EffectiveDimension}(b,c) support this interpretation by revealing a strong correlation between the average half-chain entanglement entropy $S_{E}$ of eigenstates of chaotic SU(2)-symmetric Hamiltonians and $\log_2 D_n$, and a weaker correlation with $\log_2 D_n d_n$. Hence, these figures confirm that it is the bond-algebra irreps, not the full spin-symmetry sectors, that determine the degrees of thermalization of energy eigenstates.

So far, the variable $n$ defined in \eqref{definitionn} has merely served as an alternative label to the spin sector $S$. We now show that it also represents the number of irreducible degrees of freedom required to describe the many-body wavefunction and thus offers a concrete picture of the varying degrees of complexity that generate the thermalization hierarchy \eqref{ThermalizationHierarchy}.

\subsection{Irreducible degrees of freedom}

Let us first consider the $n=0$ states, which live in the maximal $S=\frac{L}{2}$ spin sector. The formulas \eqref{DS} and \eqref{ds} yield $D_{0}=1$ and $d_{0}=L+1$. Hence, there must be $L+1$ degenerate energy eigenstates that transform as 1D irreps of the bond algebra \eqref{BondAlgebraSpins}. One of them is the ferromagnetic state
\begin{equation}
    \ket{0} =  \ket{\uparrow,\uparrow,\cdots,\uparrow}, \label{FM}
\end{equation}
which lies in the $S^{z}_{\mathrm{tot}}=\frac{L}{2}$ magnetization sector. Completely symmetric under all elements of the permutation algebra $\mathcal{A}_{\mathrm{bond}}$, it has no internal dynamics under SU(2)-symmetric Hamiltonians \eqref{Heinsenberg} and behaves as the ``vacuum''. One obtains the remaining states belonging to the trivial bond irrep by successively applying the total-spin lowering operator
\begin{align}
    \ket{n=0,m}
    &=
    \sqrt{\frac{(L-m)!}{m!\,L!}}\,
    \left(\hat{S}^{-}_{\mathrm{tot}}\right)^{m}\ket{0} \nonumber \\
    &=
    \frac{1}{\sqrt{\binom{L}{m}}}
    \sum_{j_{1}<\cdots<j_m}
    \ket{j_{1},\dots,j_{m}},
    \label{Dicke}
\end{align}
with $m=0,\dots,L$, where
$\ket{j_{1},\dots,j_{m}}
=
\hat{S}^{-}_{j_1}\cdots \hat{S}^{-}_{j_m}\ket{0}$. If $\ket{0}$ is the vacuum, then $\ket{0,m}$ contains $m$ elementary excitations (magnons) and belongs to the $S^{z}_{\mathrm{tot}}=\frac{L}{2}-m$ magnetization subspace.

Since the states \eqref{Dicke} are uniform superpositions of the basis elements $\ket{j_{1},\dots,j_{m}}$, they remain fully symmetric under permutations, so they continue to transform in the trivial irrep $(n=0)$ of the bond algebra and therefore are exact eigenstates of arbitrary SU(2)-symmetric Hamiltonians. Physically, they may be interpreted as $m$ zero-momentum magnons condensed into a dynamically frozen collective mode. This absence of collective degrees of freedom naturally connects them to the towers of QBMSs generated by restricted spectrum-generating algebras~\cite{PhysRevD.2.2944,bohm2024twenty,PhysRevB.102.085140}.

To move beyond the trivial irrep, consider generic one-magnon states,
\begin{equation}
    \ket{\psi}
    =
    \sum_{j=0}^{L-1} c_j \ket{j},
    \qquad
    \ket{j}=\hat S_j^- \ket{0},
\end{equation}
which span a subspace of dimension $L$. We may decompose them into $n=0$ and $n=1$ contributions,
\begin{align}
    \ket{\psi}&=\ket{\psi^{(0)}}+\ket{\psi^{(1)}} \nonumber \\
    &= \sum_{j}\left[c^{(0)}_j+c^{(1)}_j\right]\ket{j}.
\end{align}
The $n=0$ component is simply the average of the coefficients $c_j$,
\begin{equation}
    c_j^{(0)}
    = c^{(0)} = 
    \frac{1}{L}\sum_{\ell=1}^{L} c_\ell, \label{Scalar}
\end{equation}
so that $\ket{\psi^{(0)}}$ is proportional to \eqref{Dicke} for $m=1$.

Imposing $\langle \psi^{(0)} | \psi^{(1)}\rangle = 0$, we find that the $n=1$ contribution satisfies
\begin{equation}
    c_{j}^{(1)}=c_j-c_{j}^{(0)} \Rightarrow \sum_{j=0}^{L-1} c^{(1)}_j=0. \label{constraint1}
\end{equation}
For one-magnon states other than $\ket{\psi^{(0)}}$, the $c_j^{(1)}$ are nonzero depend irreducibly on the spatial coordinate $j$. By that, we mean that these coefficients must be represented by a rank-1 tensor or length $L$, unlike the $c^{(0)}_j$ contribution, which can be reduced to a scalar as in \eqref{Scalar}. Furthermore, this rank-1 tensor is subject to the constraint \eqref{constraint1}, and  therefore spans an $(L-1)$-dimensional subspace. We interpret this coordinate dependence as one irreducible degree of freedom (IDOF): a collective mode whose dynamical subspace has dimension $L-1$. Consistently, \eqref{DS} gives $D_{n=1}=L-1$, the dimension of the ``standard'' representation of $S_L$.

From the one-magnon state $\ket{\psi^{(1)}}$, one can obtain other states with one IDOF lying in different $m$-magnon sectors by repeatedly applying the spin-lowering operator,
\begin{align}
    \ket{n=1,m}
    &=
    \mathcal{N}_{m}
    \left(\hat{S}^{-}_{\mathrm{tot}}\right)^{m-1}
    \sum_{j}c_{j}^{(1)}\ket{j} \nonumber \\
    &=
    \mathcal{N}_{m}
    \sum_{j_1<\cdots<j_m}
    \left(
        \sum_{\alpha=1}^{m}
        c_{j_\alpha}^{(1)}
    \right)
    \ket{j_1,\dots,j_m}, \label{1DOF}
\end{align}
where $\mathcal N_m$ is a normalization constant and $m \in \{1,\dots,L-1\}$, yielding a total $L-1$ steps in the ladder and matching $d_S=L-1$ from \eqref{ds}. Crucially, each step in this ladder is still determined by the $L$-component vector $(c_0^{(1)},\dots,c_{L-1}^{(1)})$ constrained by $\sum_j c_j^{(1)}=0$. Consequently, the dynamically accessible subspace at every $m$ remains $(L-1)$-dimensional, reflecting the presence of only one IDOF.

The same structure persists in higher $n$ sectors. For example, a generic two-magnon state can be decomposed into orthogonal components with
$n\in\{0,1,2\}$,
\begin{align}
    \ket{\psi}
    &=
    \sum_{j_1<j_2}
    c_{j_1j_2}
    \ket{j_1,j_2}
    \nonumber \\
    &=
    \ket{\psi^{(0)}}
    +
    \ket{\psi^{(1)}}
    +
    \ket{\psi^{(2)}}
    \nonumber \\
    &=
    \sum_{j_1<j_2}
    \left[
    c_{j_1j_2}^{(0)}
    +
    c_{j_1j_2}^{(1)}
    +
    c_{j_1j_2}^{(2)}
    \right]
    \ket{j_1,j_2}, \label{Generic2magnon}
\end{align}
where $c_{j_1j_2}=c_{j_2j_1}$. In total, there are $\binom{L}{2}$ independent coefficients, matching the dimension of the two-magnon Hilbert space. Again, the $n=0$ component is the average
\begin{equation}
    c_{j_1j_2}^{(0)}
    =c^{(0)}=
    \frac{1}{\binom{L}{2}}
    \sum_{j_1<j_2}
    c_{j_1j_2}, \label{0con}
\end{equation}
and therefore coordinate-independent.

Equation \eqref{1DOF} with $m=2$ implies that the $n=1$ contributions to \eqref{Generic2magnon} must be additive,
\begin{equation}
    c_{j_1j_2}^{(1)}
    =
    c^{(1)}_{j_1}+c^{(1)}_{j_2},
    \qquad
    \sum_j c^{(1)}_j=0, \label{1con}
\end{equation}
which contains $L-1$ independent coefficients. In Appendix \ref{AppendixB}, we show how to extract $c^{(1)}_j$ from the row sums of $c_{ij}$ as
\begin{equation}
    c^{(1)}_j
    =
    \frac{1}{L-2}
    \sum_{i\neq j}
    \left(
        c_{ij}-c^{(0)}
    \right). \label{aconstraints}
\end{equation}

Subtracting the $n=0,1$ contributions from \eqref{Generic2magnon}, we obtain
\begin{equation}
    c^{(2)}_{j_1j_2}
    =
    c_{j_1j_2}
    -
    c^{(1)}_{j_1}
    -
    c^{(1)}_{j_2}
    -
    c^{(0)}.
\end{equation}
The remaining coefficients $c^{(2)}_{ij}$ cannot be expressed as any combination 
of single-coordinate functions, that is, they depend irreducibly on \emph{both} 
coordinates $i$ and $j$ simultaneously. We therefore interpret them as carrying 
two IDOF: a collective mode that explores a subspace parameterized by two spatial labels. This irreducibility is further 
reflected in the orthogonality conditions
\begin{equation}
    \sum_{i\neq j} c^{(2)}_{ij} = 0,
    \qquad
    \sum_{i<j} c^{(2)}_{ij} = 0,
\end{equation}
which enforce orthogonality to the $n=0$ and $n=1$ sectors respectively, and 
play the role of the two constraints analogous to \eqref{constraint1}. Removing 
these two lower sectors from the $\binom{L}{2}$-dimensional two-magnon Hilbert 
space, the two IDOF explore a subspace of dimension
\begin{equation}
    \binom{L}{2}-1-(L-1)
    =
    \binom{L}{2}-L
    =
    \frac{L(L-3)}{2},
\end{equation}
which exactly matches $D_{2}$ from \eqref{DS}.

The examples above reveal a general correspondence between the bond-irrep 
label $n = L/2 - S$ and the number of irreducible degrees of freedom 
underlying a many-body state. \textit{We define a state as carrying $n$ 
irreducible degrees of freedom (IDOF) if its wavefunction depends 
irreducibly on $n$ spatial coordinates, meaning it cannot be expressed 
as a combination of functions of fewer than $n$ independent spatial 
labels --- the precise analogue of an irreducible tensor of rank $n$.} 
Crucially, this number is independent of the number of elementary 
excitations $m$: the fully symmetric ($n=0$) states \eqref{Dicke} 
contain arbitrarily many magnons while carrying no IDOF, whereas the ``standard'' ($n=1$) states \eqref{1DOF} carry 
exactly one IDOF independently of $m$.

\begin{figure}
    \centering
    \includegraphics[width=0.8\linewidth]{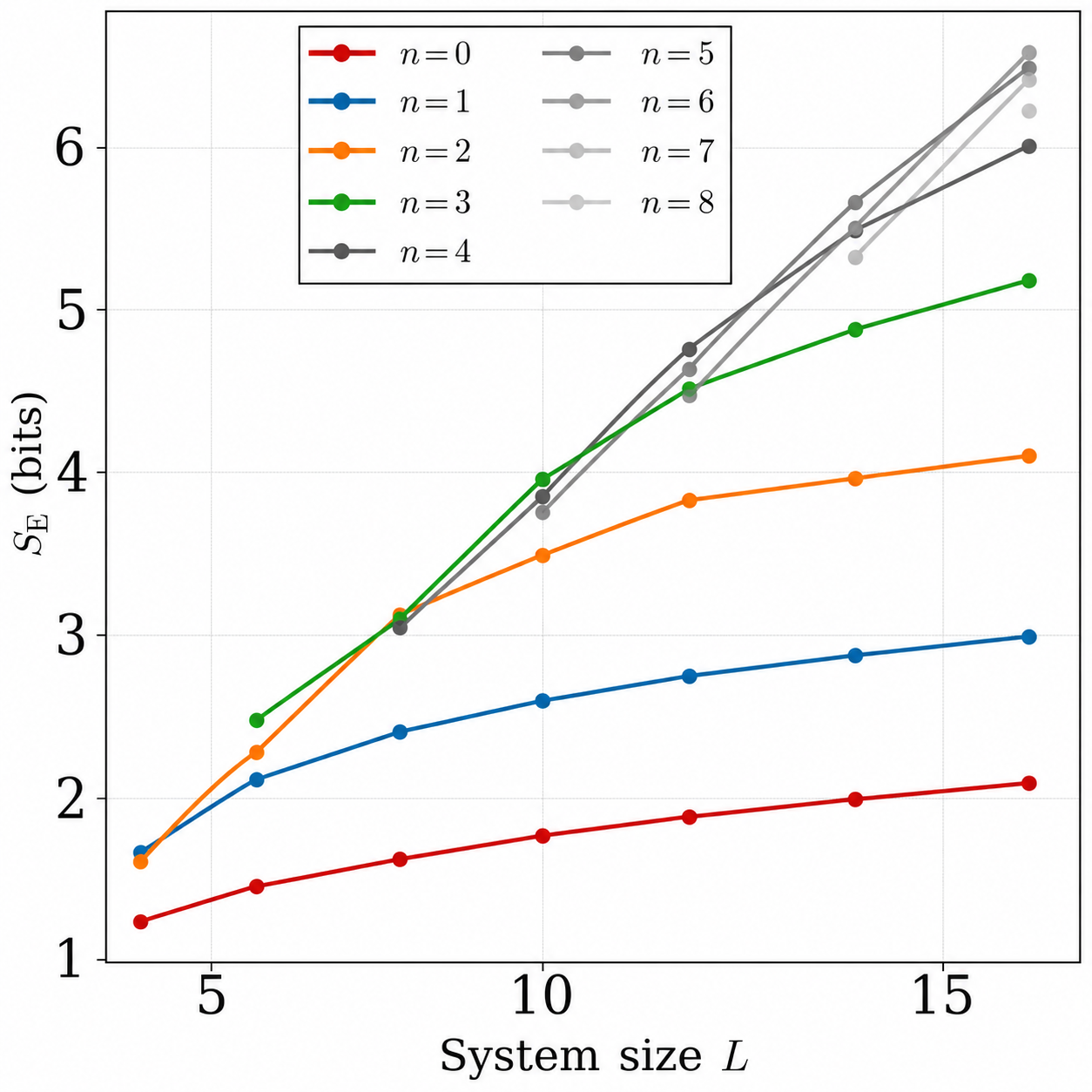}
    \caption{Scaling of the half-chain entanglement entropy $S_E$ averaged over eigenstates of the Hamiltonian \eqref{Heinsenberg} with fixed $n$ IDOF.}
    \label{fig:Hierarchy}
\end{figure}

The dimension $D_n$ of each IDOF subspace grows rapidly with $n$ and 
peaks at
\begin{equation}
    n_{*} = \frac{L-\sqrt{L+2}}{2}, \label{MostThermalSector}
\end{equation}
which scales with system size. As Fig.~\ref{fig:Hierarchy} shows, the 
entanglement entropies of the different $n$ sectors branch out progressively 
with $L$, forming an envelope of volume-law states centered around 
$n \simeq n^*$. Since $n^*$ grows with $L$, any sector at fixed $n$ occupies 
an increasingly subthermal position in the hierarchy as $L \to \infty$, 
making low-IDOF sectors natural candidates for hosting atypical eigenstates 
that evade conventional thermalization.

\section{Symmetry breaking and protected subspaces}

The existence of low-IDOF sectors alone does not produce QMBS, because exact SU(2) symmetry organizes the whole spectrum into degenerate multiplets labeled by $n=L/2-S$. To realize QMBS from a target sector $n=n_{\mathrm{target}}$, we proceed in two steps. First, we break the conventional SU(2) symmetry while preserving a dynamical SU(2) structure generated by the total-spin ladder operators. Although this lifts the multiplet degeneracies, the residual U(1) symmetry still places each ladder step within a fixed-$S^{z}_{\mathrm{tot}}$ sector. As a result, the low-IDOF states remain near the edge of the magnetization-resolved spectrum rather than embedded within its thermal bulk. Second, we break the residual U(1) symmetry while preserving the target sector $n=n_{\mathrm{target}}$. The resulting perturbation hybridizes the generic sectors but leaves the target subspace invariant, yielding a collection of nonthermal eigenstates embedded in an otherwise ergodic many-body continuum [see Fig.~\ref{HamiltonianDecomposition}(b,c)].

\subsection{Lifting degeneracies}

We now show that perturbations drawn from the commutant algebra lift degeneracies without modifying the bond irreducible spaces. As a result, they preserve the hierarchy of irreducible degrees of freedom while lifting the multiplicity structure of the spectrum.

Let $\hat X$ be a Hermitian element of the commutant algebra \eqref{CommutantAlgebra} that does not belong to the common center $\mathcal Z=\mathcal A_{\mathrm{bond}}\cap\mathcal A_{\mathrm{comm}}$.\footnote{Such an element can always be constructed: since $\mathcal A_{\mathrm{comm}}$ is closed under Hermitian conjugation, for any $\hat O\in\mathcal A_{\mathrm{comm}}$, the operator $\hat X=\frac12(\hat O+\hat O^\dagger)$ also belongs to $\mathcal A_{\mathrm{comm}}$ and satisfies $\hat X^\dagger=\hat X$.} We then consider ``shifted'' Hamiltonians of the form
\begin{equation}
\hat H_X
=
\sum_\alpha J_\alpha \hat H_\alpha
+
\mathcal{E} \hat X .
\label{ModifiedHamiltonian}
\end{equation}
The inclusion of $\hat X$ enlarges the bond algebra,
\begin{equation}
\mathcal A_{\mathrm{bond}}'
=
\mathrm{Alg}
\left(
\mathcal A_{\mathrm{bond}},
\hat X
\right),
\end{equation}
while reducing its commutant to
\begin{equation}
\mathcal A_{\mathrm{comm}}'
=
\mathcal A_{\mathrm{comm}}
\cap
\{\hat X\}' ,
\end{equation}
where $\{\hat X\}'$ denotes the set of operators commuting with $\hat X$. Nevertheless, the original decomposition \eqref{Decomposition} continues to provide a basis in which both $\hat H$ and $\hat X$ are block diagonal. In particular,
\begin{equation}
\hat H
=
\bigoplus_\lambda
\left(
H_\lambda
\otimes
\mathbb I_{d_\lambda}
\right),
\qquad
\hat X
=
\bigoplus_\lambda
\left(
\mathbb I_{D_\lambda}
\otimes
X_\lambda
\right),
\end{equation}
where $D_\lambda=\dim\mathcal H_\lambda^{\mathrm{bond}}$, $d_\lambda=\dim\mathcal H_\lambda^{\mathrm{comm}}$, and $X_\lambda$ are $d_\lambda\times d_\lambda$ Hermitian matrices acting on $\mathcal H_\lambda^{\mathrm{comm}}$. It immediately follows that
\begin{equation}
\hat H_X
=
\bigoplus_\lambda
\left(
H_\lambda\otimes\mathbb I_{d_\lambda}
+
\mathcal{E} \ 
\mathbb I_{D_\lambda}
\otimes
X_\lambda
\right).
\label{BlockHprime}
\end{equation}

Since each $X_\lambda$ is Hermitian, it admits a spectral decomposition
\begin{equation}
X_\lambda
=
\sum_\mu
x_{\lambda,\mu}
P_{\lambda,\mu},
\end{equation}
where $P_{\lambda,\mu}$ are the orthogonal projectors onto the eigenspaces of $X_\lambda$, satisfying $\sum_\mu P_{\lambda,\mu}=\mathbb I_{d_\lambda}$. Substituting this decomposition into \eqref{BlockHprime} yields
\begin{equation}
\hat H_X
=
\bigoplus_\lambda
\sum_\mu
\left(
H_\lambda
+
\mathcal{E} \ x_{\lambda,\mu}
\mathbb I_{D_\lambda}
\right)
\otimes
P_{\lambda,\mu}.
\end{equation}

Thus, the perturbation by $\hat X$ preserves the decomposition into $\lambda$ sectors while lifting the $d_\lambda$-dimensional multiplicity spaces according to the spectrum of $X_\lambda$. Equivalently, the original decomposition \eqref{Decomposition} is refined into
\begin{equation}
\mathcal H
=
\bigoplus_{\lambda,\mu}
\left(
\mathcal H_\lambda^{\mathrm{bond}}
\otimes
\mathcal H_{\lambda,\mu}^{\mathrm{comm}}
\right),
\end{equation}
where $\mathcal H_{\lambda,\mu}^{\mathrm{comm}}=P_{\lambda,\mu}\mathcal H_\lambda^{\mathrm{comm}}$ denotes the eigenspace of $X_\lambda$ with eigenvalue $x_{\lambda,\mu}$.

Going back to the Heisenberg example~\eqref{Heinsenberg}, let us add a uniform magnetic field in the $z$ direction,
\begin{equation}
\hat H_{\mathrm{SU(2)}_d}
=
\hat H_{\mathrm{SU(2)}}
-
B\hat S^z_{\mathrm{tot}} .
\label{DynSU2Hamiltonian}
\end{equation}
Then
\begin{equation}
[\hat H_{\mathrm{SU(2)}_d},\hat S^\pm_{\mathrm{tot}}]
=
\mp B\hat S^\pm_{\mathrm{tot}},
\qquad
[\hat H_{\mathrm{SU(2)}_d},\hat S^z_{\mathrm{tot}}]
=
0 .
\end{equation}
Thus, the addition of the magnetic field reduces the commutant algebra to the subalgebra generated by $\hat S^z_{\mathrm{tot}}$. The operators $\hat S^\pm_{\mathrm{tot}}$ no longer commute with $\hat H_{\mathrm{SU(2)}_d}$, but instead act as spectrum-generating operators, shifting energies by $\mp B$. In this sense, the original SU(2) commutant is converted into a dynamical SU(2) symmetry or spectrum-generating algebra (SGA)~\cite{buvca2019non,PhysRevB.102.041117,PhysRevD.2.2944,PhysRevB.102.085140,tang2022multimagnon,bohm2024twenty}.

\subsection{Protected subspaces and quantum many-body scars}

The dynamical SU(2) structure lifts the exact multiplet degeneracies, but all IDOF sectors remain organized into ladders connected by $\hat S^\pm_{\mathrm{tot}}$. We now show how additional perturbations can selectively preserve chosen bond irreps while thermalizing the rest of the spectrum. The resulting protected subspaces provide a natural setting for a hierarchy of QMBS and may be viewed as a generalization of restricted spectrum-generating algebras from individual states to entire dynamically isolated subspaces.

Let us now take a Hamiltonian of the form \eqref{DecompositionHamiltonian} or its shifted version \eqref{ModifiedHamiltonian} and modify it as
\begin{equation} 
\hat{H}_V = \hat{H} + \hat{V}. 
\end{equation}
We choose $\hat V$ so that it breaks the commutant structure of $\hat H$ on the full Hilbert space while preserving a target set of bond irreps. Specifically, for $\lambda\in\Lambda_{\mathrm{target}}\subset\Lambda$, we require
\begin{equation}
\hat V
\ket{w_{\lambda,\alpha,\beta}}
=
\sum_{\beta'=1}^{d_\lambda}
v_{\lambda}^{\beta'\beta}
\ket{w_{\lambda,\alpha,\beta'}} ,
\label{ProtectedCondition}
\end{equation}
where $\ket{w_{\lambda,\alpha,\beta}}$ denotes the basis \eqref{SpecialBasis} adapted to the double-commutant decomposition \eqref{Decomposition}. Equivalently,
\begin{equation}
\left.
\hat V
\right|_{\mathcal H_\lambda}
=
\mathbb I_{D_\lambda}
\otimes
V^{\mathrm{comm}}_\lambda,
\qquad
\lambda\in\Lambda_{\mathrm{target}},
\label{RestrictedDCTCondition}
\end{equation}
where $V^{\mathrm{comm}}_\lambda$ acts on the multiplicity space $\mathcal H^{\mathrm{comm}}_\lambda$. Thus, $\hat V$ may act nontrivially within the commutant spaces of the target sectors, but it leaves the bond irreducible spaces $\mathcal H^{\mathrm{bond}}_\lambda$ unchanged.

As a result, the Hilbert space reorganizes as
\begin{align}
    \mathcal{H}&= \mathcal{H}_{\mathrm{target}} \oplus \mathcal{H}_{\mathrm{thermal}},
\end{align}
where
\begin{equation}
    \mathcal{H}_{\mathrm{target}} = \bigoplus_{\lambda \in \Lambda_{\mathrm{target}}} \left(\mathcal{H}^{\mathrm{bond}}_{\lambda} \otimes \mathcal{H}^{\mathrm{comm}}_{\lambda}\right).
\end{equation}
The target bond irreps remain dynamically isolated, whereas the remaining sectors hybridize into a large thermal subspace. This realizes a restricted double-commutant structure: the original decomposition survives only on $\mathcal H_{\mathrm{target}}$, where it protects nonthermal eigenstates embedded in an otherwise ergodic spectrum.

\subsubsection{Local example}

As an example, let us modify the Hamiltonian \eqref{DynSU2Hamiltonian} as
\begin{equation}
    \hat{H} = \hat{H}_{\mathrm{SU(2)}_d} + \kappa \ \hat{O}_{\pi} \hat{S}^{x}_{\mathrm{tot}}\hat{O}_{\pi}, \label{HamiltonianOpi}
\end{equation}
where $\kappa \in \mathbb{R}$, $\hat{S}^{x}_{\mathrm{tot}}$ breaks the residual U(1) symmetry, and
\begin{equation}
    \hat{O}_{\pi} = \sum_{j=0}^{L-1}(-1)^j\hat{\bm{S}}_{j}\cdot\hat{\bm{S}}_{j+1}. \label{Opi}
\end{equation}
Since the states $\ket{0,m}$ defined in \eqref{Dicke} are fully permutation-symmetric, every nearest-neighbor bond operator
$\hat{\bm S}_j\cdot\hat{\bm S}_{j+1}$
acts identically on them. Because
$\sum_j (-1)^j=0$
for even $L$, the alternating bond operator \eqref{Opi} annihilates the entire $n=0$ sector. We thus obtain
\begin{align}
    &(\mathrm{i}) \ \hat{H}\ket{0,m} = (E_0 -m \ B )\ket{0,m} \nonumber \\
    & \ \ \ \  \text{for} \  m \in \{0,1,\dots,L\}  \nonumber, \\
    & (\mathrm{ii}) \ \hat{S}^{-}_{\mathrm{tot}}\ket{0,L} = 0 \label{RSGA0}
\end{align}
which consists of a restricted spectrum-generating algebra (RSGA) of order $L+1$~\cite{PhysRevB.102.085140}. Therefore, even though U(1) has been broken globally, \eqref{Dicke} are still exact energy eigenstates, or 1D irreps of a \emph{restricted} bond algebra.

The operator \eqref{Opi} also annihilates
\begin{equation}
\ket{1,m,\pi} = (\hat{S}^{-}_{\mathrm{tot}})^{m-1}\sum_{j}(-1)^{j}\hat{S}^{-}_{j}\ket{0}.
\end{equation}
These states span the momentum-$\pi$ component of the $n=1$ sector~\eqref{1DOF}. Therefore, if \eqref{HamiltonianOpi} has translation invariance, we obtain
\begin{align}
    &(\mathrm{i}) \ \hat{H}\ket{n=1,m,\pi} = (E_\pi -m \ B )\ket{1,m,\pi} \nonumber \\
    & \ \ \ \  \text{for} \  m \in \{1,\dots,L-1\}  \nonumber, \\
    & (\mathrm{ii}) \ \hat{S}^{-}_{\mathrm{tot}}\ket{1,L-1,\pi} = 0 \label{RSGA1}
\end{align}
which defines a RSGA of order $L-1$.

Fig.~\ref{HamiltonianDecomposition}(b) shows the half-chain entanglement entropy $S_E$ versus energy for the eigenstates of a translation-invariant Hamiltonian of the form \eqref{HamiltonianOpi}. The towers \eqref{RSGA0} and \eqref{RSGA1} appear as atypical states embedded within an otherwise chaotic spectrum. The $n=0$ tower (red), previously identified in Ref.~\cite{PhysRevX.14.041069}, exhibits exceptionally low entanglement and therefore constitutes a strong family of QMBS. In contrast, the $n=1$ tower (blue) possesses higher, but still subthermal, entanglement and realizes a weaker family of scars. This distinction is consistent with the thermalization hierarchy of Fig.~\ref{fig:Hierarchy}, where increasing $n$ corresponds to larger effective dynamical dimensions $D_n$ and stronger ergodic mixing.

In the previous example, only the $q=\pi$ component of the $n=1$ bond irrep was protected. More generally, one may construct local operators $\hat O_q$ that preserve other momentum sectors $q\neq 0$. However, the resulting scars rely on translation symmetry and therefore do not realize the full restricted double-commutant mechanism \eqref{RestrictedDCTCondition}. To demonstrate that this mechanism is completely general, we now give a nonlocal construction that protects the entire $n=1$ bond irrep.

\subsubsection{Non-local example}

Consider the Hamiltonian
\begin{equation}
    \hat H
    =
    \hat H_{\mathrm{SU(2)}_d}
    +
    \kappa\,
    \hat O_1
    \hat S^x_{\mathrm{tot}}
    \hat O_1, \label{HNonlocal}
\end{equation}
where
\begin{equation}
    \hat O_1
    =
    \hat{\bm S}_{\mathrm{tot}}^2
    -
    S_1(S_1+1)\,\mathbb I,
    \qquad
    S_1=\frac{L}{2}-1.
    \label{NonlocalHamiltonian}
\end{equation}
Since every state in the $n=1$ sector satisfies
\begin{equation}
    \hat{\bm S}_{\mathrm{tot}}^2
    \ket{1,m}
    =
    S_1(S_1+1)
    \ket{1,m},
\end{equation}
the operator $\hat O_1$ annihilates the entire $n=1$ bond irrep. Consequently,
\begin{equation}
    \hat O_1
    \hat S^x_{\mathrm{tot}}
    \hat O_1
    \ket{1,m}
    =
    0,
\end{equation}
and all states $\ket{1,m}$ remain in the $n=1$ irrep of $S_L$.

Here, the $n=0$ sector is not annihilated by $\hat O_1$. However, because $D_0=1$, the perturbation acts only on the commutant multiplicity space and leaves the bond-irrep structure unchanged, precisely as required by \eqref{RestrictedDCTCondition}. As a result, the corresponding states remain weakly entangled and continue to occupy the lowest level of the thermalization hierarchy, as shown in Fig.~\ref{HamiltonianDecomposition}(c).

More importantly, the nonlocal construction protects the entire $n=1$ bond irrep rather than a single momentum sector. Since $D_1=L-1$, the perturbed Hamiltonian preserves $L-1$ independent one-IDOF modes. Each generates a ladder of $L-1$ states under the action of $\hat S^-_{\mathrm{tot}}$, producing $L-1$ distinct scar towers embedded in the many-body spectrum.

\section{Discussion }

In this work, we showed that the decomposition of the Hilbert space of a many-body quantum system into the irreducible representations of its bond and commutant algebras establishes a thermalization hierarchy. Dynamically isolated subspaces correspond to irreducible representations of the bond algebra, whose dimensions $D_\lambda$ strongly correlate with local entanglement and control the degree of thermalization within a given sector. In this picture, $\log D_\lambda$ plays the role of an effective entropy associated with the dynamical subspace accessible to a given bond irrep.

Using spin-$1/2$ chains as an example where the double-commutant decomposition is particularly transparent, we showed that the bond-irrep subspaces admit a natural interpretation as the configuration spaces explored by ``irreducible degrees of freedom'' (IDOF). Concretely, a state with $n$ IDOF is specified by a tensor of order $n$ transforming irreducibly under the bond algebra. Moreover, such collective degrees of freedom do not correspond to elementary excitations (magnons), but instead emerge from the structure of the many-body Hamiltonian.

Finally, we showed that, by breaking the symmetries of the Hamiltonian in all but a few target sectors, one obtains a restricted double-commutant decomposition that organizes the target sectors into towers of IDOF subspaces embedded within a thermal spectrum. This prescription generalizes the notion of restricted spectrum-generating algebras by protecting entire dynamical subspaces rather than individual eigenstates, thereby establishing a hierarchy of quantum many-body scars with varying degrees of subthermality and robustness against symmetry breaking.

Crucially, our framework does not fundamentally rely on SU(2) symmetry or Schur--Weyl duality, but only on the existence of a nontrivial double-commutant structure. Therefore, analogous thermalization hierarchies may arise in systems with SU($N$) symmetries, Temperley--Lieb algebras, $\eta$-pairing Hubbard models, or kinetically constrained systems, where the physical interpretation of the IDOF and their constrained dynamics may be richer.

\begin{acknowledgments}
We thank Haylen Gerhard, Daniel Sussman, Andre G. Fonseca and Sachin Vaidya for useful discussions. P.F.C and W.A.B. are thankful for the support of the startup funds from Emory University.
\end{acknowledgments}

\emph{Data availability.} All data is available upon reasonable request.

\bibliography{poster}

\newpage

\appendix

\section{Examples of the double-commutant decomposition}\label{AppendixDecomposition}

Let $\mathcal{H}=(\mathbb{C}^{2})^{\otimes L}$ be the Hilbert space of $L$
spin-$1/2$ degrees of freedom, and let $\mathcal{L}(\mathcal{H})$ denote the
algebra of linear operators acting on $\mathcal{H}$. The associative algebras
$\mathcal{A}_{S_L}$ and $\mathcal{A}_{\mathfrak{su}(2)}$, generated by the
actions of the symmetric group $S_L$ (permutations of spins) and the total spin
operators, respectively form a double commutant pair (that is, they are mutual centralizers of each other) in
$\mathcal{L}(\mathcal{H})$, i.e.,
\begin{equation}
    \mathcal{A}_{S_L}' = \mathcal{A}_{\mathfrak{su}(2)}, 
    \qquad
    \mathcal{A}_{\mathfrak{su}(2)}' = \mathcal{A}_{S_L}.
\end{equation}
The corresponding double-commutant decomposition of the Hilbert space takes the
form of the Schur--Weyl duality~\cite{fulton2013representation,stevens2016schur},
\begin{equation}
    \mathcal{H}
    =
    \bigoplus_{\lambda}
    \mathcal{H}^{S_L}_{\lambda}
    \otimes
    \mathcal{H}^{\mathfrak{su}(2)}_{\lambda}.
    \label{SchurWeylAppendix}
\end{equation}
Here each sector $\mathcal{H}^{S_L}_{\lambda}
    \otimes
    \mathcal{H}^{\mathfrak{su}(2)}_{\lambda}$ of dimension $d_\lambda D_{\lambda}$ corresponds to a
pair of irreducible representations: the symmetric group $S_L$ acts
nontrivially only on $\mathcal{H}^{S_L}_{\lambda}$, where it realizes a
$D_\lambda$-dimensional irrep, while the total spin algebra acts nontrivially only on
$\mathcal{H}^{\mathfrak{su}(2)}_{\lambda}$, where it realizes a $d_\lambda$-dimensional irrep.

Concretely, for every $\mathcal{H}^{S_L}_{\lambda}
    \otimes
    \mathcal{H}^{\mathfrak{su}(2)}_{\lambda}$ sector there is an orthonormal basis
\begin{equation}
    \{\ket{w_{\lambda,\alpha,\beta}}=\ket{u_{\lambda,\alpha}} \otimes \ket{v_{\lambda,\beta}}\}, \label{DecompositionBasis}
\end{equation}
where $\ket{u_{\lambda,\alpha}} \in \mathcal{H}_{\lambda}^{S_L}$ with $\alpha \in \{1,\dots,D_{\lambda}\}$, and $\ket{v_{\lambda,\beta}} \in \mathcal{H}^{\mathfrak{su}(2)}_{\lambda}$ with $\beta \in \{1,\dots,d_{\lambda}\}$. In that basis, any operator $\hat{O}_{S_L} \in \mathcal{A}_{S_L}$ acts as
\begin{equation}
    \hat{O}_{S_L}\ket{u_{\lambda,\alpha}} \otimes \ket{v_{\lambda,\beta}} = \sum_{\alpha'}M^{\lambda}_{\alpha\alpha'}(\hat{O}_{S_L})\ket{u_{\lambda,\alpha'}} \otimes \ket{v_{\lambda,\beta}},
\end{equation}
where $M^{\lambda}(\hat{O}_{S_L})$ is the $D_{\lambda} \times D_{\lambda}$ matrix representation of $\hat{O}_{S_L}$ restricted to $\mathcal{H}_{\lambda}^{S_L}$. Similarly, for any $\hat{O}_{\mathfrak{su}(2)} \in \mathcal{A}_{\mathfrak{su}(2)}$, we have
\begin{equation}
    \hat{O}_{\mathfrak{su}(2)}\ket{u_{\lambda,\alpha}} \otimes \ket{v_{\lambda,\beta}} = \sum_{\beta'}\ket{u_{\lambda,\alpha}} \otimes N^{\lambda}_{\beta\beta'}(\hat{O}_{\mathfrak{su}(2)})\ket{v_{\lambda,\beta'}},
\end{equation}
where $N^{\lambda}(\hat{O}_{\mathfrak{su}(2)})$ is the $d_{\lambda} \times d_{\lambda}$ matrix representation of $\hat{O}_{\mathfrak{su}(2)}$ restricted to $\mathcal{H}_{\lambda}^{\mathfrak{su}(2)}$. Summing over all irreps $\lambda$, we obtain the following resolution of the identity in the full Hilbert space $\mathcal{H}$,
\begin{equation}
    \sum_{\lambda}\sum_{\alpha=1}^{D_{\lambda}}\sum_{\beta=1}^{d_{\lambda}} \ket{w_{\lambda,\alpha,\beta}}\bra{w_{\lambda,\alpha,\beta}} = \mathbb{I}_{\mathcal{H}}
\end{equation}
In that basis, the matrix elements of $\hat{O}_{S_L} \in \mathcal{A}_{S_L}$ are
\begin{equation}
    \bra{w_{\lambda,\alpha,\beta}}\hat{O}_{S_L}\ket{w_{\lambda',\alpha',\beta'}} = \delta_{\lambda\lambda'}M^{\lambda}_{\alpha\alpha'}(\hat{O}_{S_L})\delta_{\beta\beta'}, \label{BondMatrixElements}
\end{equation}
while the matrix elements of $\hat{O}_{\mathfrak{su}(2)} \in \mathcal{A}_{\mathfrak{su}(2)}$ read
\begin{equation}
    \bra{w_{\lambda,\alpha,\beta}}\hat{O}_{{\mathfrak{su}(2)}}\ket{w_{\lambda',\alpha',\beta'}} = \delta_{\lambda\lambda'}\delta_{\alpha\alpha'}N^{\lambda}_{\beta\beta'}(\hat{O}_{{\mathfrak{su}(2)}}).
\end{equation}
This construction makes manifest that $\mathcal{A}_{S_L}$ and
$\mathcal{A}_{\mathfrak{su}(2)}$ act irreducibly on complementary tensor
factors within each $\lambda$ sector, while leaving the other factor invariant. Let us now see explicitly how it works for small system sizes.

\subsection{Two spins}

\begin{table}[h]
\centering
\begin{tabular}{c|cc}
 & $\{e\}$ & $\{(01)\}$ \\
\hline
$|C|$ & 1 & 1 \\
\hline
trivial & 1 & 1 \\
sign & 1 & -1 \\
\end{tabular}
\caption{Character table of $S_2$ acting on $\{0,1\}$, with class sizes.}
\label{Character2}
\end{table}

The Hilbert space of two spin-$1/2$ degrees of freedom is four-dimensional and
is spanned by the basis vectors
\begin{equation}
    B = \{\ket{\uparrow\uparrow}, \ket{\uparrow\downarrow},
    \ket{\downarrow\uparrow},\ket{\downarrow\downarrow}\},
    \label{LocalBasis}
\end{equation}
where $\ket{\uparrow}$ and $\ket{\downarrow}$ are the basis vectors of the
single-spin Hilbert space, satisfying
$\hat{S}_{i}^{z}\ket{\uparrow} = \frac{\hbar}{2}\ket{\uparrow}$ and
$\hat{S}_{i}^{z}\ket{\downarrow} = -\frac{\hbar}{2}\ket{\downarrow}$, for the site coordinate
$i=0,1$.

Since we only have two sites, the permutation algebra $\mathcal{A}_{S_2}$ is
generated by the symmetric group $S_2$, which contains two elements: the identity
and the pairwise permutation $\hat{\pi}_{01}$. The latter acts on the basis
\eqref{LocalBasis} as $\hat{\pi}_{01}\ket{\sigma_0\sigma_1}
    =
    \ket{\sigma_1\sigma_0}$ for $
    \sigma_i \in \{\uparrow,\downarrow\}$, or explicitly
\begin{align}
    \hat{\pi}_{01}\ket{\uparrow\uparrow} &= \ket{\uparrow\uparrow},\nonumber \\
    \hat{\pi}_{01}\ket{\uparrow\downarrow} &= \ket{\downarrow\uparrow}, \nonumber \\
    \hat{\pi}_{01}\ket{\downarrow\uparrow} &= \ket{\uparrow\downarrow}, \nonumber \\
    \hat{\pi}_{01}\ket{\downarrow\downarrow} &= \ket{\downarrow\downarrow}.
\end{align}
Therefore, the matrix representation of $\hat{\pi}_{01}$ in the basis \eqref{LocalBasis} is
\begin{equation}
    [\hat{\pi}_{01}]_{B} = \begin{pmatrix}
        1 & 0 & 0 & 0 \\
        0 & 0 & 1 & 0 \\
        0 & 1 & 0 & 0 \\
        0 & 0 & 0 & 1
    \end{pmatrix}.
\end{equation}
The eigenvectors of $\hat{\pi}_{01}$ are
\begin{align}
    &\text{eigenvalue} \ +1: \begin{cases}
        \ket{w_{\lambda=\mathrm{triv},\alpha=1,\beta=1}} = \ket{\uparrow\uparrow} \\
        \ket{w_{\mathrm{triv},1,2}} = \frac{1}{\sqrt{2}}(\ket{\uparrow\downarrow}+\ket{\downarrow\uparrow}) \\
        \ket{w_{\mathrm{triv},1,3}} = \ket{\downarrow\downarrow} 
    \end{cases}
    \nonumber \\
        &\text{eigenvalue} \ -1: \ket{w_{\textrm{sgn},1,1}} = \frac{1}{\sqrt{2}}(\ket{\uparrow\downarrow}-\ket{\downarrow\uparrow}).
\end{align}
They all define 1D irreps of $\mathcal{A}_{S_2}$: $\ket{w_{\mathrm{triv},1,1}}, \ket{w_{\mathrm{triv},1,2}}$ and $\ket{w_{\mathrm{triv},1,3}}$ transform according to the trivial ($\lambda=\mathrm{triv}$) irrep, while $\ket{w_{\mathrm{sgn},1,1}}$ transforms according to the ``sign'' ($\lambda=\mathrm{sgn}$) irrep. In the language of the decomposition \eqref{SchurWeylAppendix}, $\mathcal{H}_{\mathrm{triv}}^{S_2}$ and $\mathcal{H}_{\mathrm{sgn}}^{S_2}$ have dimensions $D_{\mathrm{triv}}=D_{\mathrm{sgn}}=1$.

What about the $\mathcal{H}_{\mathrm{triv}}^{\mathfrak{su}(2)}$ and $\mathcal{H}_{\mathrm{sgn}}^{\mathfrak{su}(2)}$ subspaces? The generators of $\mathcal{A}_{\mathfrak{su}(2)}$ are $\{\hat{S}_{\mathrm{tot}}^{+},\hat{S}_{\mathrm{tot}}^{-},\hat{S}_{\mathrm{tot}}^{2}\}$. The only simultaneous eigenvector of all three generators is the apin-0 singlet $\ket{w_{\mathrm{sgn},1,1}}$, with eigenvalue 0. It therefore defines a 1D irrep of $\mathcal{A}_{\mathfrak{su}(2)}$ whose subspace $\mathcal{H}_{0}^{\mathfrak{su}(2)}$ has dimension $d_{\mathrm{sgn}}=d_{0}=1$. 

Meanwhile, applying the generators to any state in the symmetric irrep eventually produces all three states. For example,
\begin{equation}
    \hat{S}^{-}_{\mathrm{tot}}\ket{w_{\mathrm{triv},1,1}} = \ket{w_{\mathrm{triv},1,2}}, \ \hat{S}^{-}_{\mathrm{tot}}\ket{w_{\mathrm{triv},1,2}}=\ket{w_{\mathrm{triv},1,3}}.
\end{equation}
Therefore, they define a spin-1 irrep and a subspace $\mathcal{H}_{\mathrm{1}}^{\mathfrak{su}(2)}$ of dimension $d_{\mathrm{triv}}=d_{1}=3$.

Finally, the Schur-Weyl decomposition for the $L=2$ case is
\begin{equation}
    \mathcal{H} = (\mathcal{H}^{S_2}_{\mathrm{triv}} \otimes \mathcal{H}_{1}^{\mathfrak{su}(2)}) \oplus (\mathcal{H}^{S_2}_{\mathrm{sgn}} \otimes \mathcal{H}_{0}^{\mathfrak{su}(2)}),
\end{equation}
where $D_{\mathrm{triv}}=1$, $d_{1}=3$, $D_{\mathrm{sgn}}=1$, $d_{0}=1$, yielding $D_{\mathrm{triv}}d_{1}+D_{\mathrm{sgn}}d_{0}=4$, as expected from the character table of $S_2$ (see Tab. \ref{Character2}).

\subsection{Three spins}

\begin{table}[h]
\centering
\begin{tabular}{c|ccc}
 & $\{e\}$ & $\{(12),(13),(23)\}$ & $\{(123),(132)\}$ \\
\hline
$|C|$ & 1 & 3 & 2 \\
\hline
trivial      & 1 & 1  & 1 \\
standard    & 2 & 0  & -1 \\
sign  & 1 & -1 & 1 \\
\end{tabular}
\caption{Character table of $S_3$.}
\label{CharcacterTableS3}
\end{table}

The Hilbert space of three spin-1/2 degrees of freedom is eight-dimensional and spanned by
\begin{align}
    B = &\{\ket{\uparrow\uparrow\uparrow},\nonumber \\
    & \ \ket{\uparrow\uparrow\downarrow},\ket{\uparrow\downarrow\uparrow},\ket{\downarrow\uparrow\uparrow}, \nonumber \\
    & \ \ket{\uparrow\downarrow\downarrow},\ket{\downarrow\uparrow\downarrow}, \ket{\downarrow\downarrow\uparrow}, \nonumber \\
    & \ \ket{\downarrow\downarrow\downarrow}\}. \label{Basis3spins}
\end{align}

The permutation algebra $\mathcal{A}_{S_3}$ is now generated by $S_3$, which has $6$ elements: the identity, three pairwise permutations $\hat{\pi}_{01}$, $\hat{\pi}_{02}$, and $\hat{\pi}_{12}$, as well as two cyclic permutations defined as
\begin{align}
    \hat{\pi}_{012} = \hat{\pi}_{01}\hat{\pi}_{12} , \nonumber \\
    \hat{\pi}_{021} = \hat{\pi}_{12}\hat{\pi}_{01}.
\end{align}
Our notation for cyclic permutations follows the pattern
\begin{align}
    &\hat{\pi}_{012}: \quad 0 \to 1 \to 2 \to 0, \nonumber \\
    &\hat{\pi}_{021}: \quad 0 \to 2 \to 1 \to 0,
\end{align}
such that
\begin{align}
    \hat{\pi}_{012}\ket{\sigma_0\sigma_1\sigma_2}&=\ket{\sigma_2\sigma_0\sigma_1}, \nonumber \\
    \hat{\pi}_{021}\ket{\sigma_0\sigma_1\sigma_2}&=\ket{\sigma_1\sigma_2\sigma_0}.
\end{align}

The eight-dimensional Hilbert space of must be spanned by the irreps of
\begin{equation}
    S_3 = \{\mathbb{I}_{\mathcal{H}}, \hat{\pi}_{01},\hat{\pi}_{02},\hat{\pi}_{12},\hat{\pi}_{012},\hat{\pi}_{021}\},
\end{equation}
each irrep having its multiplicity. We now show that, with the basis \eqref{Basis3spins}, we can form a $D_{\lambda}=1$ irrep of multiplicity $d_{\lambda}=4$ and a $D_{\lambda}=2$ irrep with multiplicity $d_{\lambda}=2$, adding up to $8$ dimensions.

To find the 1D irrep, we notice that the states
\begin{align}
    \ket{w_{\mathrm{triv},1,1}} &= \ket{\uparrow\uparrow\uparrow}, \nonumber \\
    \ket{w_{\mathrm{triv},1,2}} &= \frac{1}{\sqrt{3}}(\ket{\uparrow\uparrow\downarrow}+\ket{\uparrow\downarrow\uparrow}+\ket{\downarrow\uparrow\uparrow}), \nonumber \\
    \ket{w_{\mathrm{triv},1,3}} &= \frac{1}{\sqrt{3}}(\ket{\uparrow\downarrow\downarrow}+\ket{\downarrow\uparrow\downarrow}+\ket{\downarrow\downarrow\uparrow}), \nonumber \\
    \ket{w_{\mathrm{triv},1,4}} &= \ket{\downarrow\downarrow\downarrow}, \label{FullySymmetric}
\end{align}
are all simultaneous eigenvectors of every element of $S_3$ with eigenvalue +1. They all transform according to the trivial $(\lambda=\mathrm{triv})$ representation of $\mathcal{A}_{S_3}$. Their 4-fold multiplicity comes from the fact that they form a spin-3/2 quadruplet by the repeated action of $\hat{S}_{\mathrm{tot}}^{\pm}$.

The other 1D irrep of $S_{3}$ is the ``sign'' irrep $(\lambda=-)$. However, it is impossible to realize such an irrep for 3 spin-1/2 degrees of freedom. To see how, notice that one may completely antisymmetrize an element $\ket{\sigma_{0}\sigma_{1}\sigma_{2}}$ of the basis \eqref{Basis3spins} as
\begin{equation}
    \ket{\psi_{\mathrm{anti}}} = \sum_{\pi \in S_3} \mathrm{sgn}(\pi) \ket{\sigma_{\pi(0)}\sigma_{\pi(1)}\sigma_{\pi(2)}}, \label{antisymmetrization}
\end{equation}
where the sign $\mathrm{sgn}(\pi)$ of the permutation $\pi$ is $(-1)^{s(\pi)}$, where $s(\pi)$ is the number of pairwise ``swaps'' necessary to implement $\pi$. We list the signs of the elements of $S_3$ in Table \ref{Tab:1}.

\begin{table}[h]
\centering
\begin{tabular}{c c c c}
\hline
\textbf{Operator} & \textbf{Cycle form} & \textbf{Swaps needed} & \textbf{Sign} \\
\hline
$\mathbb{I}_{\mathcal{H}}$ & $(0)(1)(2)$ & 0 & $+1$ \\
$\hat{\pi}_{01}$ & $(0\,1)$ & 1 & $-1$ \\
$\hat{\pi}_{02}$ & $(0\,2)$ & 1 & $-1$ \\
$\hat{\pi}_{12}$ & $(1\,2)$ & 1 & $-1$ \\
$\hat{\pi}_{012}$ & $(0\,1\,2)$ & 2 & $+1$ \\
$\hat{\pi}_{021}$ & $(0\,2\,1)$ & 2 & $+1$ \\
\hline
\end{tabular} 
\caption{Elements of $S_3$ written in terms of permutation operators acting on sites $0,1,2$.}
\label{Tab:1}
\end{table}

For example, applying \eqref{antisymmetrization}  to $\ket{\uparrow\uparrow\downarrow}$ yields
\begin{align}
    \ket{\psi_{\mathrm{anti}}} &= \ket{\uparrow\uparrow\downarrow}-\ket{\uparrow\uparrow\downarrow}-\ket{\downarrow\uparrow\uparrow} \nonumber \\
    \ & - \ket{\uparrow\downarrow\uparrow} + \ket{\downarrow\uparrow\uparrow} + \ket{\uparrow\downarrow\uparrow} = 0,
\end{align}
and similarly of all elements of \eqref{Basis3spins}. Since any vector of the Hilbert space is a linear combination os those basis elements, we conclude that the only completely antisymmetric vector under permutations is the null vector. The same conclusion follows for all $L \geq 3$. This means there we cannot realize the ``sign'' representation of $S_L$ with $L \geq 3$ in the Hilbert space of spin-1/2 degrees of freedom.

The first higher dimensional irrep of $S_{3}$ is the ``standard'' representation of dimension $D_{\mathrm{std}}=L-1=2$, as one can read from the character table \ref{CharcacterTableS3}. A simple basis of the corresponding subspaces are the simultaneous eigenstates of the cyclic elements $\hat{\pi}_{012}$ and $\hat{\pi}_{021}$. Since $(\hat{\pi}_{012})^3=(\hat{\pi}_{021})^3=\mathbb{I}_{\mathcal{H}}$, their eigenvalues are $\{1,e^{\frac{2\pi\mathrm{i}}{3}},e^{-\frac{2\pi\mathrm{i}}{3}}\}$. Eigenvalue $1$ corresponds to the fully symmetric states \eqref{FullySymmetric}. Notice that
\begin{align}
     &\hat{\pi}_{012}\left[\ket{\uparrow\uparrow\downarrow}+e^{-\mathrm{i}\frac{2\pi}{3}}\ket{\uparrow\downarrow\uparrow}+e^{\mathrm{i}\frac{2\pi}{3}}\ket{\downarrow\uparrow\uparrow}\right] \nonumber \\
     &= \ket{\downarrow\uparrow\uparrow}+e^{-\mathrm{i}\frac{2\pi}{3}}\ket{\uparrow\uparrow\downarrow}+e^{\mathrm{i}\frac{2\pi}{3}}\ket{\uparrow\downarrow\uparrow} \nonumber \\
     &= e^{-\mathrm{i}\frac{2\pi}{3}}\left[\ket{\uparrow\uparrow\downarrow}+e^{-\mathrm{i}\frac{2\pi}{3}}\ket{\uparrow\downarrow\uparrow}+e^{\mathrm{i}\frac{2\pi}{3}}\ket{\downarrow\uparrow\uparrow}\right],
\end{align}
and
\begin{align}
     &\hat{\pi}_{012}\left[\ket{\downarrow\downarrow\uparrow}+e^{-\mathrm{i}\frac{2\pi}{3}}\ket{\downarrow\uparrow\downarrow}+e^{\mathrm{i}\frac{2\pi}{3}}\ket{\uparrow\downarrow\downarrow}\right] \nonumber \\
     &= \ket{\uparrow\downarrow\downarrow}+e^{-\mathrm{i}\frac{2\pi}{3}}\ket{\downarrow\downarrow\uparrow}+e^{\mathrm{i}\frac{2\pi}{3}}\ket{\downarrow\uparrow\downarrow} \nonumber \\
     &= e^{-\mathrm{i}\frac{2\pi}{3}}\left[\ket{\downarrow\downarrow\uparrow}+e^{-\mathrm{i}\frac{2\pi}{3}}\ket{\downarrow\uparrow\downarrow}+e^{\mathrm{i}\frac{2\pi}{3}}\ket{\uparrow\downarrow\downarrow}\right].
\end{align}
Therefore, the states
\begin{align}
    \ket{w_{\mathrm{std},1,1}}&=\frac{1}{\sqrt{3}}\left(\ket{\uparrow\uparrow\downarrow}+e^{-\mathrm{i}\frac{2\pi}{3}}\ket{\uparrow\downarrow\uparrow}+e^{\mathrm{i}\frac{2\pi}{3}}\ket{\downarrow\uparrow\uparrow}\right), \nonumber \\
    \ket{w_{\mathrm{std},1,2}}&=\frac{1}{\sqrt{3}}\left(\ket{\downarrow\downarrow\uparrow}+e^{-\mathrm{i}\frac{2\pi}{3}}\ket{\downarrow\uparrow\downarrow}+e^{\mathrm{i}\frac{2\pi}{3}}\ket{\uparrow\downarrow\downarrow}\right), \label{SpinDoublet1}
\end{align}
are eigenvectors of $\hat{\pi}_{012}$ with eigenvalue $e^{\mathrm{i}\frac{2\pi}{3}}$ and eigenvectors of $\hat{\pi}_{021}$ with eigenvalue $e^{-\mathrm{i}\frac{2\pi}{3}}$. Similarly, the states
\begin{align}
    \ket{w_{\mathrm{std},2,1}}&=\frac{1}{\sqrt{3}}\left(\ket{\uparrow\uparrow\downarrow}+e^{\mathrm{i}\frac{2\pi}{3}}\ket{\uparrow\downarrow\uparrow}+e^{-\mathrm{i}\frac{2\pi}{3}}\ket{\downarrow\uparrow\uparrow}\right), \nonumber \\
    \ket{w_{\mathrm{std},2,2}}&=\frac{1}{\sqrt{3}}\left(\ket{\downarrow\downarrow\uparrow}+e^{\mathrm{i}\frac{2\pi}{3}}\ket{\downarrow\uparrow\downarrow}+e^{-\mathrm{i}\frac{2\pi}{3}}\ket{\uparrow\downarrow\downarrow}\right), \label{SpinDoublet2}
\end{align}
are eigenvectors of $\hat{\pi}_{021}$ with eigenvalue $e^{-\mathrm{i}\frac{2\pi}{3}}$ and eigenvectors of $\hat{\pi}_{021}$ with eigenvalue $e^{\mathrm{i}\frac{2\pi}{3}}$.

Now let us see how these four states transform under the pairwise permutators. We have
\begin{align}
    \hat{\pi}_{01}\ket{w_{\mathrm{std},1,1}} &= \frac{1}{\sqrt{3}}\left(\ket{\uparrow\uparrow\downarrow}+e^{\mathrm{i}\frac{2\pi}{3}}\ket{\uparrow\downarrow\uparrow}+e^{-\mathrm{i}\frac{2\pi}{3}}\ket{\downarrow\uparrow\uparrow}\right) \nonumber \\
    &= \ket{w_{\mathrm{std},2,1}}, \nonumber \\
    \hat{\pi}_{12}\ket{w_{\mathrm{std},1,1}} &= \frac{1}{\sqrt{3}}\left(\ket{\uparrow\downarrow\uparrow}+e^{-\mathrm{i}\frac{2\pi}{3}}\ket{\uparrow\uparrow\downarrow}+e^{\mathrm{i}\frac{2\pi}{3}}\ket{\downarrow\uparrow\uparrow}\right) \nonumber \\
    &=e^{-\mathrm{i}\frac{2\pi}{3}}\ket{w_{\mathrm{std},2,1}}, \nonumber \\
    \hat{\pi}_{02} \ket{w_{\mathrm{std},1,1}}&= \frac{1}{\sqrt{3}}\left(\ket{\downarrow\uparrow\uparrow}+e^{-\mathrm{i}\frac{2\pi}{3}}\ket{\uparrow\downarrow\uparrow}+e^{\mathrm{i}\frac{2\pi}{3}}\ket{\uparrow\uparrow\downarrow}\right) \nonumber \\
    &= e^{\mathrm{i}\frac{2\pi}{3}}\ket{w_{\mathrm{std},2,1}}. \label{Permutations1}
\end{align}
Moreover,
\begin{align}
    \hat{\pi}_{01}\ket{w_{\mathrm{std},1,2}} &= \frac{1}{\sqrt{3}}\left(\ket{\downarrow\downarrow\uparrow}+e^{\mathrm{i}\frac{2\pi}{3}}\ket{\downarrow\uparrow\downarrow}+e^{-\mathrm{i}\frac{2\pi}{3}}\ket{\uparrow\downarrow\downarrow}\right) \nonumber \\
    &= \ket{w_{\mathrm{std},2,2}}, \nonumber \\
    \hat{\pi}_{12}\ket{w_{\mathrm{std},1,2}} &= \frac{1}{\sqrt{3}}\left(\ket{\downarrow\uparrow\downarrow}+e^{-\mathrm{i}\frac{2\pi}{3}}\ket{\downarrow\downarrow\uparrow}+e^{\mathrm{i}\frac{2\pi}{3}}\ket{\uparrow\downarrow\downarrow}\right) \nonumber \\
    &= e^{-\mathrm{i}\frac{2\pi}{3}}\ket{w_{\mathrm{std},2,2}}, \nonumber \\
    \hat{\pi}_{02}\ket{w_{\mathrm{std},1,2}} &= \frac{1}{\sqrt{3}}\left(\ket{\uparrow\downarrow\downarrow}+e^{-\mathrm{i}\frac{2\pi}{3}}\ket{\downarrow\uparrow\downarrow}+e^{\mathrm{i}\frac{2\pi}{3}}\ket{\downarrow\downarrow\uparrow}\right) \nonumber \\
    &= e^{\mathrm{i}\frac{2\pi}{3}}\ket{w_{\mathrm{std},2,2}}. \label{Permutations2}
\end{align}

The results \eqref{Permutations1} and \eqref{Permutations2} show that $S_3$ maps $\ket{w_{\mathrm{std}_{1,1}}} \leftrightarrow \ket{w_{\mathrm{std},2,1}}$ and $\ket{w_{\mathrm{std}_{1,2}}} \leftrightarrow \ket{w_{\mathrm{std},2,2}}$, splitting those four states into two doublets. Hence, we have a two-dimensional standard representation $(D_{\mathrm{std}}=2)$ with multiplicity $d_{\mathrm{std}}=2$.

Applying the lowering operator $\hat{S}^{-}_{\mathrm{tot}}$, we see that the states in \eqref{SpinDoublet1} form a spin-1/2 doublet,
\begin{align}
    \hat{S}^{-}_{\mathrm{tot}}\ket{w_{\mathrm{std},1,1}} &= \frac{1}{3}\big[\ket{\downarrow\uparrow\downarrow} +\ket{\uparrow\downarrow\downarrow} + e^{-\mathrm{i}\frac{2\pi}{3}}(\ket{\downarrow\downarrow\uparrow}+\ket{\uparrow\downarrow\downarrow}) \nonumber \\
    & \ +e^{\mathrm{i}\frac{2\pi}{3}}(\ket{\downarrow\downarrow\uparrow}+\ket{\downarrow\uparrow\downarrow})\big] \nonumber \\
    &= \frac{2\cos(\frac{2\pi}{3})}{3}\big[\ket{\downarrow\downarrow\uparrow}+e^{-\mathrm{i}\frac{2\pi}{3}}\ket{\downarrow\uparrow\downarrow}+e^{\mathrm{i}\frac{2\pi}{3}}\ket{\uparrow\downarrow\downarrow}\big] \nonumber \\
    &= -\frac{1}{\sqrt{3}}\ket{w_{\mathrm{std},1,2}}.
\end{align}
Similarly, the states in \eqref{SpinDoublet2} also form a spin-1/2 doublet,
\begin{align}
    \hat{S}^{-}_{\mathrm{tot}}\ket{w_{\mathrm{std},2,1}} &= \frac{1}{3}\big[\ket{\downarrow\uparrow\downarrow} +\ket{\uparrow\downarrow\downarrow} + e^{\mathrm{i}\frac{2\pi}{3}}(\ket{\downarrow\downarrow\uparrow}+\ket{\uparrow\downarrow\downarrow}) \nonumber \\
    & \ +e^{-\mathrm{i}\frac{2\pi}{3}}(\ket{\downarrow\downarrow\uparrow}+\ket{\downarrow\uparrow\downarrow})\big] \nonumber \\
    &= \frac{2\cos(\frac{2\pi}{3})}{3}\big[\ket{\downarrow\downarrow\uparrow}+e^{\mathrm{i}\frac{2\pi}{3}}\ket{\downarrow\uparrow\downarrow}+e^{-\mathrm{i}\frac{2\pi}{3}}\ket{\uparrow\downarrow\downarrow}\big] \nonumber \\
    &= -\frac{1}{\sqrt{3}}\ket{w_{\mathrm{std},2,2}},
\end{align}
confirming we have $d_{\mathrm{std}}=d_{1/2}=2$. 

Therefore, the decomposition \eqref{SchurWeylAppendix} for $L=3$ is
\begin{equation}
    \mathcal{H} = (\mathcal{H}^{S_3}_{\mathrm{triv}} \otimes \mathcal{H}^{\mathfrak{su}(2)}_{3/2} )\oplus (\mathcal{H}^{S_3}_{\mathrm{std}} \otimes \mathcal{H}^{\mathfrak{su}(2)}_{1/2}), 
\end{equation}
where $D_{\mathrm{triv}}=1$, $d_{3/2}=4$, $D_{\mathrm{std}}=2$, $d_{1/2}=2$, and $D_{\mathrm{triv}}d_{3/2}+D_{\mathrm{std}}d_{1/2}=8$.

\section{Projection onto the one-coordinate sector in the two-magnon Hilbert space}
\label{AppendixB}

In this Appendix, we derive Eq.~\eqref{aconstraints}.  We work in the two-magnon Hilbert space, whose wavefunctions are specified by symmetric coefficients
$c_{ij}=c_{ji}$ with $i\neq j$.  Equivalently, the independent amplitudes are
$c_{ij}$ with $i<j$.  The natural inner product between two such coefficient arrays is
\begin{equation}
    \langle c |d \rangle
    =
    \sum_{i<j} c_{ij}^{*} d_{ij}.
\end{equation}
We want to decompose a generic two-magnon coefficient array into three mutually orthogonal pieces,
\begin{equation}
    c_{ij}
    =
    c_{ij}^{(0)}
    +
    c_{ij}^{(1)}
    +
    c_{ij}^{(2)} ,
\end{equation}
where the $n=0$ component is constant, the $n=1$ component is additive in the two coordinates, and the remaining part belongs to the $n=2$ sector.

The $n=0$ component is the projection onto the constant vector in the two-magnon Hilbert space.  Therefore
\begin{equation}
    c_{ij}^{(0)}
    =
    c^{(0)}
    =
    \frac{1}{\binom{L}{2}}
    \sum_{i<j} c_{ij}.
\end{equation}
Subtracting this component, define
\begin{equation}
    \widetilde c_{ij}
    =
    c_{ij}-c^{(0)}.
\end{equation}
By construction,
\begin{equation}
    \sum_{i<j}\widetilde c_{ij}=0.
\end{equation}

The $n=1$ subspace consists of additive two-magnon amplitudes of the form
\begin{equation}
    c_{ij}^{(1)}
    =
    a_i+a_j,
    \qquad
    \sum_i a_i=0.
    \label{appendix:additive}
\end{equation}
The constraint $\sum_i a_i=0$ removes the constant component, since a uniform shift
$a_i\mapsto a_i+\alpha$ only produces another constant contribution
$c_{ij}^{(1)}\mapsto c_{ij}^{(1)}+2\alpha$.  Hence the additive sector contains
$L-1$ independent coefficients, as expected for the $n=1$ sector.

To extract the coefficients $a_j$ from a generic two-magnon state, consider the row sums of the zero-average coefficient array,
\begin{equation}
    R_j
    =
    \sum_{i\neq j} \widetilde c_{ij}
    =
    \sum_{i\neq j}
    \left(
        c_{ij}-c^{(0)}
    \right).
    \label{appendix:rowsum}
\end{equation}
If $\widetilde c_{ij}$ were purely additive, $\widetilde c_{ij}=a_i+a_j$, then
\begin{align}
    R_j
    &=
    \sum_{i\neq j} (a_i+a_j)
    \nonumber\\
    &=
    \sum_{i\neq j} a_i
    +
    (L-1)a_j
    \nonumber\\
    &=
    -a_j+(L-1)a_j
    \nonumber\\
    &=
    (L-2)a_j,
\end{align}
where we used $\sum_i a_i=0$.  Therefore the additive coefficients are
\begin{equation}
    a_j
    =
    \frac{1}{L-2}
    \sum_{i\neq j}
    \left(
        c_{ij}-c^{(0)}
    \right).
    \label{appendix:aj}
\end{equation}
Identifying $a_j\equiv c_j^{(1)}$ gives Eq.~\eqref{aconstraints}.

It remains to show that this prescription indeed gives the orthogonal projection onto the $n=1$ subspace.  Define
\begin{equation}
    c_{ij}^{(1)}
    =
    a_i+a_j,
\end{equation}
with $a_j$ given by Eq.~\eqref{appendix:aj}, and define the residual
\begin{equation}
    c_{ij}^{(2)}
    =
    c_{ij}
    -
    c^{(0)}
    -
    a_i
    -
    a_j.
    \label{appendix:c2def}
\end{equation}
The row sum of the residual is
\begin{align}
    \sum_{i\neq j} c_{ij}^{(2)}
    &=
    \sum_{i\neq j}
    \left(
        c_{ij}-c^{(0)}
    \right)
    -
    \sum_{i\neq j}(a_i+a_j)
    \nonumber\\
    &=
    R_j-(L-2)a_j
    \nonumber\\
    &=
    0.
    \label{appendix:c2rowsum}
\end{align}
Thus the residual is characterized by vanishing row sums.

The three pieces are mutually orthogonal.  First, the constant component is orthogonal to the additive component because
\begin{equation}
    \sum_{i<j}(a_i+a_j)
    =
    (L-1)\sum_i a_i
    =
    0.
\end{equation}
Second, the constant component is orthogonal to the residual because
\begin{equation}
    \sum_{i<j} c_{ij}^{(2)}
    =
    \frac{1}{2}
    \sum_j
    \sum_{i\neq j} c_{ij}^{(2)}
    =
    0.
\end{equation}
Finally, the additive component is orthogonal to the residual since
\begin{align}
    \sum_{i<j} (a_i+a_j)^{*} c_{ij}^{(2)}
    &=
    \sum_j a_j^{*}
    \sum_{i\neq j} c_{ij}^{(2)}
    \nonumber\\
    &=
    0.
\end{align}
Therefore the decomposition
\begin{equation}
    c_{ij}
    =
    c^{(0)}
    +
    a_i+a_j
    +
    c_{ij}^{(2)}
\end{equation}
is an orthogonal decomposition of the two-magnon Hilbert space into the
$n=0$, $n=1$, and $n=2$ sectors.

The dimension counting is also consistent.  The full two-magnon Hilbert space has dimension
\begin{equation}
    \binom{L}{2}.
\end{equation}
The constant sector has dimension $1$, the additive sector has dimension $L-1$, and the residual sector has dimension
\begin{equation}
    \binom{L}{2}-1-(L-1)
    =
    \binom{L}{2}-L
    =
    \frac{L(L-3)}{2}.
\end{equation}
This is precisely the expected dimension of the $n=2$ sector in the two-magnon subspace.

%\bibliographystyle{apsrev4-2}
% Bibliography

\end{document}